\documentclass[epj]{svjour}
\usepackage{graphicx}
\usepackage{amssymb}

\newcommand\derr{\partial_r}
\newcommand\dereta{\partial_\eta}
\newcommand\dert{\partial_\mathrm{t}}
\newcommand\CCUN{{\cal C}^{(+)}}
\newcommand\CCD{{\cal C}^{(-)}}
\newcommand\CCK{{\cal C}^{(k)}}
\newcommand\IOP{{\cal I}}
\newcommand\RL{L}
\newcommand\KKK{{\cal K}}
\newcommand\INVPM{ I_\pm}
\newcommand\INVUN{ I_+}
\newcommand\INVD{I_-}
\newcommand\INVK{I_k}
\newcommand\DELTAX{\Delta{x_0}}
\newcommand\ZETA{\Delta{x_0}}
\newcommand\RHOST{\rho_\mathrm{st}}

\Roman{table}

\begin{document}
\title{Non-linear effects and shock formation\\
in the focusing of a spherical acoustic wave}
\subtitle{Numerical simulations and experiments in liquid helium}
\author{C. Appert \inst{1}, C. Tenaud \inst{2}, X. Chavanne \inst{1}, S.Balibar
\inst{1},
F. Caupin \inst{1} and D.
d'Humi\`eres \inst{1}}

\institute{Laboratoire de Physique Statistique de l'ENS, associ\'e au CNRS
et aux Universit\'es Paris 6 et 7, 24 rue Lhomond 75231 Paris Cedex 05,
France
  \and  LIMSI, Batiment 508, Universit\'e Paris-Sud, 91405 Orsay, France}
\mail{appert@lps.ens.fr}
\date{draft: \today}
\abstract{The focusing of acoustic waves is used to study nucleation
phenomena in liquids. At large amplitude, non-linear effects are important
so that the magnitude of pressure or density oscillations is difficult to
predict. We present a calculation of these oscillations in a spherical
geometry.
 We show that the main source of non-linearities is the
shape of the equation of state of the liquid,
enhanced by the spherical
geometry.
We also show that the formation of shocks cannot be ignored
beyond a certain oscillation amplitude.
The shock length is estimated by an analytic calculation
based on the characteristics method.
In our numerical simulations, we have treated the shocks
with a WENO scheme.
We obtain a very good agreement
with experimental measurements which were recently performed in liquid helium.
The comparison between numerical and experimental
results allows in particular to calibrate the vibration
of the ceramics used to produce the wave,
as a function of the applied voltage.
\PACS{
       {6740.-w}{}   \and
       {4325.+y}{}   \and
       {6260.+v}{}
      } 
} 
\authorrunning{C. Appert et al.}
\titlerunning{Nonlinear focusing of a spherical acoustic wave}
\maketitle

\section{Introduction}

Recent experiments have shown that acoustic waves can be used to study the
nucleation of phase transitions far from equilibrium under very clean
conditions~\cite{lambare98,caupin_b01,chavanne_b_c01}. Thanks to hemispherical piezo-electric
transducers, we have focused 1MHz acoustic waves in liquid helium
and produced large pressure and density oscillations. These waves are
quasi-spherical and, at the acoustic focus (the center), their amplitude
can be very large. We used an optical method to detect
  the nucleation
of bubbles by the negative swings of the waves~\cite{lambare98,caupin_b01}. This nucleation
occurs beyond a certain threshold in the sound amplitude which needs to be
  determined as accurately as possible, in order to
compare with independent theoretical predictions.
We later obtained
evidence for the nucleation of crystals by the positive swings~\cite{chavanne_b_c01} and had the
same need.

In the absence of non-linear effects, the measurement of the
nucleation threshold
would
be simple to do. For example, one could study the nucleation
  as a function of the static pressure in the experimental
cell, and then use a linear extrapolation~\cite{caupin_b01}.
  However, we expect non-linear effects to occur, especially
in cavitation studies. Indeed, the homogeneous nucleation of bubbles
occurs near the ``spinodal limit'' where the compressibility diverges and
the sound velocity vanishes. When an acoustic wave is produced in a fluid,
with an amplitude such that during the negative swings the sound velocity
approaches zero, it is clear that the wave must be highly distorted.
Non-linear effects have been already noticed by several
authors~\cite{sirotyuk,roy_m_a90,nemirovskii90}.

We thus try to calculate the non-linear focusing of the acoustic waves.
We start with the spherical geometry, because
in a first approximation, everything depends only on
the radial distance $r$ from the center. As we shall
see (section \ref{sect_theory}),
this calculation still appears difficult because
the focusing of acoustic waves leads to the formation of
shocks at all amplitudes in a spherical geometry,
and their treatment is not quite straightforward.
We first obtain this
result and the associated shock length
from an analytic calculation which uses the methods of
characteristics (section \ref{sect_charact}).
Our calculation extends the former work of
Nemirovskii~\cite{nemirovskii90} to the spherical case, except that we neglect
the coupling with heat modes.
It is done in the spirit of Greenspan and Nadim's work~\cite{greenspan_n93},
though in our case it is slightly more tricky due to the
shape of the equation of state. We make it quantitative by
using the equation of state of liquid helium~\cite{maris91} which is well
established.
For weak oscillation amplitudes of the transducer, shocks
can be ignored and the pressure calculated at the focal point by
simulating the Euler equations using a finite difference method (section \ref{sect_simul}).
Indeed, shocks form with an infinitesimal amplitude at a distance r
from the center which is much less than our mesh size, so that one can neglect
them.
At larger amplitude, shocks cannot be ignored.
In order to treat the shocks,
we have adapted to the case of helium a code devoted
to shock simulation, based on a WENO scheme
(section \ref{sect_weno}).
In the end, we obtain the amplitude of the density oscillation at
the focus as a
function of an important parameter, the amplitude of the displacement
  of the transducer surface
where waves are generated. This transducer is a piezo-electric ceramic.
We choose 1MHz for the frequency of the waves, in order
to compare with the experiments.
In parallel, we have built an experiment to measure
the focusing in a quasi-spherical
  geometry. As explained in section \ref{sect_exp}, the results of this experiment
  allow a precise comparison with our theoretical and numerical
work~\cite{chavanne02,appert01a}. We find that the
  shape of the acoustic wave is indeed distorted at high amplitude and
very well described by our calculations, thus validating
 our theoretical method.
The
final comparison with our calculation allows us to calibrate
the efficiency of the ceramics.
As described in our
  conclusion, this work should now be extended to different geometries. One of
  them is the hemispherical geometry where, according to other experimental
  results~\cite{caupin_b01}, non-linear effects are apparently less important,
an observation
  which needs to be understood and compared with future calculations.

\section{Theory}

\label{sect_theory}

\subsection{Description of the model}

\label{sect_model}

Throughout this paper, we consider a spherical geometry. We take it as
one-dimensional since the pressure and density fields only depend on
the radial distance $r$.
We neglected dissipation since our main goal
was to compare with experiments in superfluid helium 4 which has
zero viscosity and where the attenuation of sound vanishes in the
low temperature limit.

In the case of liquid helium 4 at zero temperature, the equation
of state has been obtained by three different methods (sound
velocity extrapolations, density functional calculations, and
Monte Carlo simulations) with similar results. Maris~\cite{maris91} 
uses the simple form to relate the pressure
($P$) to the density ($\rho$):
\begin{equation}
P - P_\mathrm{sp} = \frac{b^2}{27}(\rho - \rho_\mathrm{sp})^3
\label{eos}
\end{equation}
with
\begin{eqnarray}
P_\mathrm{sp} & = & -9.6435 \,\mathrm{bar} \nonumber\\
\rho_\mathrm{sp} & = & 94.18 \,\mathrm{kg\,m^{-3}} \nonumber\\
b & = & 14.030 \:\mathrm{m^4\,s^{-1}\,kg^{-1}}. \nonumber
\end{eqnarray}
$P_\mathrm{sp}$ is the spinodal limit where the compressibility diverges,
the sound velocity vanishes and the liquid becomes totally
unstable against the formation of the vapor. At $P=0$, the density
is $\rho_0 = 145.13 \,\mathrm{kg\,m^{-3}}$. The sound speed is
then $c_\mathrm{s0} = 238.3 \,\mathrm{m\,s^{-1}}$ and the wavelength
$\lambda_0 = c_\mathrm{s0} T = 0.238 \,\mathrm{mm}$ for 1 MHz waves whose period is
$T = 1/f = 1\,\mathrm{\mu s}$.

We also considered helium 3, a lighter liquid which is not
superfluid except at very low temperature - i.e. below the
achievable temperature in our experiment. The same form
is used for the equation of state, now with
\begin{eqnarray}
P_\mathrm{sp} & = & -3.1534 \,\mathrm{bar} \nonumber\\
\rho_\mathrm{sp} & = & 53.50 \,\mathrm{kg\,m^{-3}} \nonumber\\
b & = & 19.262 \,\mathrm{m^4\,s^{-1}\,kg^{-1}}.\nonumber
\end{eqnarray}
The value of $P_\mathrm{sp}$ is less negative, which means that the inner
cohesion of liquid helium 3 is weaker than for helium 4. At $P=0$,
the density of liquid helium 3
is $\rho_0 = 81.916 \,\mathrm{kg\,m^{-3}}$. The sound speed is
then $c_\mathrm{s0} = 182.5 \,\mathrm{m\,s^{-1}}$ and the wavelength
$\lambda_0 = c_\mathrm{s0} T = 0.182 \,\mathrm{mm}$.

At the temperatures considered, the viscosity is very weak and can
be neglected though, in case of helium 3, neglecting dissipation is
an approximation which would need to be better justified.
Therefore, the numerical approximation
considers the Euler equations. In order to use a dimensionless form of the
equations, we have chosen as a time scale the period of the wave $T$,
as a length scale the wavelength at zero pressure
$\lambda_0 = c_\mathrm{s0} T$ and as a density scale the spinodal
density $\rho_\mathrm{sp}$. If we now consider $\rho$ as the dimensionless
density and $u$ as the dimensionless velocity of the fluid, the
Euler equations are written as
follows:
\begin{eqnarray}
\dert \rho + u \derr \rho + \rho \derr u
  & = & \frac{-\theta \rho u}{r}\nonumber\\
\dert u + u \derr u & = & -\frac{1}{\rho} \derr P,
\label{eqeuler}
\end{eqnarray}
or, by using the conservative variables $\rho$ and
$j = \rho u$,
\begin{eqnarray}
\dert \rho + \derr j
  & = & \frac{-\theta j}{r}\nonumber\\
\dert j + \derr [j^2/\rho + P(\rho) ] & = & \frac{-\theta j^2}{\rho r}
\label{eqeulerj}
\end{eqnarray}
where $\theta$ is respectively $0$, $1$, or $2$ in planar,
axisymetric cylindrical, or spherical geometry.
The equation of state, governing the dimensionless pressure variations, reads:
\begin{equation}
P - {P_\mathrm{sp} \over \rho_\mathrm{sp} c_\mathrm{s0}^2}=
\frac{{C}_\mathrm{o}^2}{3}(\rho - 1)^3 \label{eos_adim}
\end{equation}
with $C_\mathrm{o} = 1.848$ for helium~4 and $C_\mathrm{o} = 1.883$ for helium~3.
Note that, in order to have simple notations, we use
the same names for physical and reduced variables.
In general, calculations will be performed with reduced
variables, while numerical results and experimental
parameters will be given as physical quantities.

Boundary conditions are imposed at the center ($r=0$)
\begin{equation}
u=0,
\end{equation}
and on the transducer surface ($r=\RL_0$)
\begin{equation}
u(t) = - \omega \, \Delta x_0
\, \sin (\omega t)
\, \left[1 - \exp(-t/1.5)\right].
\end{equation}
Indeed, the motion of the transducer surface is
much smaller than a mesh size in all our simulations,
and it is sufficient
to impose an oscillating velocity on the ceramics.
The exponential term represents the response time of the
transducer. In the experiment the response time is rather equal
to $8$ $\mu$s. Here we took it shorter to have a more rapid
convergence of the calculation to the steady regime.
This has no effect on the final result, as our simulations
will always be used in the stationary regime.

In the whole paper, all the calculations are
based on Euler equations, for which shock waves do occur.
We are aware that, in superfluids as helium~4, there is no shock
wave strictly speaking.
Actually, when a steep gradient appears, it is regularized
by dispersion instead of dissipation.
This means that modes propagate with a different velocity
depending on their frequency, and thus highest frequency
modes are expelled from the steep region.
However, we assume that, as shocks form only in the focal
region, and during a limited time, not too much momentum
is lost locally due to dispersion,
and that in a first approximation, 
a viscous or dispersive regularization of the shocks
is equivalent
(note that any numerical scheme able to handle shocks
will always introduce either a dissipation or a dispersion
term in order to make shocks regular, even if Euler equations
do not contain any viscous term).
Besides, as we shall see (section \ref{sect_focal}),
we do not need to describe
the shock structure exactly, as we are above all interested
in the relaxation part of the wave.
We have found from our simulations that the negative
pressure swing only depends very weakly on the numerical
viscous regularization we use - and thus on the shock
amplitude near the focal point.
Still, it would of course be interesting
to be able to quantify further the effects
of dispersion, and this could be the object of
a further work.

Throughout the paper, we shall consider helium~4 unless it is
explicitly specified that it is helium~3. In the next sections, we
show that shock waves can occur in this system, and we compute the
radius (denoted as shock length in the followings) at which a
shock wave occurs for various oscillation amplitudes $\Delta x_0$
of the transducer surface.

\subsection{The method of characteristics}

\label{sect_charact}

In the study of compressible fluids,
the method of characteristics is a standard one~\cite{characteristics}.
It has already been used for helium in
a planar geometry (see for example~\cite{nemirovskii90}).

Here we are interested in the shock length in spherical geometry,
for the equation of state (\ref{eos}).
In this paper we define the shock length
as the distance {\em from the center} where
the shock forms.
We shall use the method of characteristics to predict a lower
bound for the shock length.

\subsubsection{Rieman invariants and characteristics}
\label{sec_rieman}

\begin{figure}[ttt]
\centerline{\includegraphics[width=7.5cm]{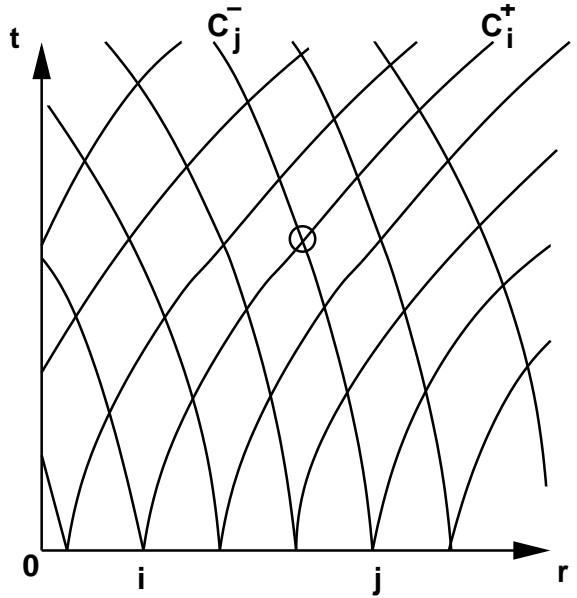}}
\caption{The two characteristic families defined in
the $(r,t)$ plane. Here each curve is labeled by the
$r$-coordinate of its intersection with the $t=0$ axis.
The point indicated by a small circle can be referred
to either by its coordinates (r,t), or by the labels
of the characteristics which intersect at this point
(i,j).
}
\label{fig_excar}
\end{figure}

  Let us first recall the principle of the method.
Solving the Euler equations means knowing
the density $\rho$ and the velocity $u$ everywhere
within a certain domain of the (r,t) plane.
In our case, the ``characteristics'' are two families of curves
$\left\{\CCUN_i\right\}_{i \in R}$ and $\left\{\CCD_i\right\}_{i \in R}$
parameterized by $i$,
where the parameter $i$ could for example be defined as the $r$-coordinate
at which the characteristic cuts the axis $t=0$
(see Fig.~\ref{fig_excar} for an illustration).
Thus, ${\cal C}^{(+)}_i$ (resp. ${\cal C}^{(-)}_i$) refers to a
single curve in the (r,t) plane, which belongs to the $(+)$
family (resp. $(-)$), and has in our case a positive slope
(resp. negative).
Each family completely covers the (r,t) plane
as $i$ is varied.
The two families of curves intersect each other.
Then, instead of locating a point in the plane by
its coordinates (r,t), it is equivalent to indicate
which particular characteristics $\CCUN_i$ and $\CCD_j$
intersect each other at this point. The point (r,t)
will then equivalently be referred to as point $(i,j)$.

These families are chosen so that,
for each family $\left\{{\cal C}^{(k)}_i\right\}$ with $k=\pm$,
there exists a quantity $\INVK$
called ``Rieman invariant'', which is a
function of $\rho$ and $u$, and obeys
a simple evolution equation
along any characteristic $\CCK_i$ of the family.
If the value of $\INVK$ is known
at one point of a characteristic $\CCK_i$ (for example
at the initial time), then it is easy to compute
it on the whole curve.

As each point of the (r,t) plane is the intersection
of two characteristics $\CCUN_i$ and $\CCD_j$,
we know the values of $\INVUN(\rho, u)$ and $\INVD(\rho, u)$
at this point. Then the density $\rho$ and the velocity $u$
are entirely determined.

Now the precise form of the Rieman invariants and corresponding
characteristics have to be derived
from the Euler equations (\ref{eqeuler})
and the equation of state for helium (\ref{eos_adim}).
The detailed calculations are given in Appendix~A.
Here we just give the results concerning
the shape of the characteristics and the Rieman invariants.
By definition, at any point, the slope of
a characteristic
  $\CCUN_i$ with equation $r = x(t)$ is $dx/dt = u + c_\mathrm{s}$
where $u$ and $c_\mathrm{s}$ are taken in $(x(t),t)$.
For a characteristic $\CCD_i$, it is $dx/dt = u - c_\mathrm{s}$.
The sound speed $c_\mathrm{s}$ is given from the equation of state
(\ref{eos_adim}) by
\begin{equation}
c_\mathrm{s} = C_\mathrm{o} (\rho-1).
\end{equation}
Then the derivative along the characteristic reads
\begin{equation}
\frac{d}{dt} = \dert + (u \pm c_\mathrm{s}) \derr. \nonumber
\end{equation}
The Rieman invariants are found to be
\begin{equation}
\INVPM \equiv C_\mathrm{o}(\rho - \ln \rho) \pm u
\label{defIk}
\end{equation}

The equations verified by the ``invariants'' $\INVK$ with $k=\pm$ read
\begin{equation}
\frac{d}{dt}\left[ \INVPM \right]
+ \theta \frac{C_\mathrm{o}(\rho -1) u}{r} = 0
\label{eq_char}
\end{equation}
where the derivative $\frac{d}{dt}$
is taken along a characteristic $\CCK_i$.
Again, $\theta$ is respectively $0$, $1$, or $2$ in planar,
cylindrical, or spherical geometry.

In the case of planar geometry
($\theta = 0$), $\INVK$ is a true invariant, since it is constant
along the corresponding characteristic,
hence its name.
In spherical or cylindrical geometries, it is not
constant, due to the source term
in  Eq.(\ref{eq_char}),
though it is still called an ``invariant''.

All this is valid at least as long as the characteristics
belonging to the {\em same} family do not intersect each other.
When they do, the corresponding $\INVK$ becomes multivalued.
This is an indication that a shock has formed,
and beyond the corresponding time, the description used in this section
breaks down.

\subsubsection{Lower bound for the shock length - Analytic calculation}

\label{sect_analytic}

\begin{figure}[ttt]
\centerline{\includegraphics[width=8.0cm]{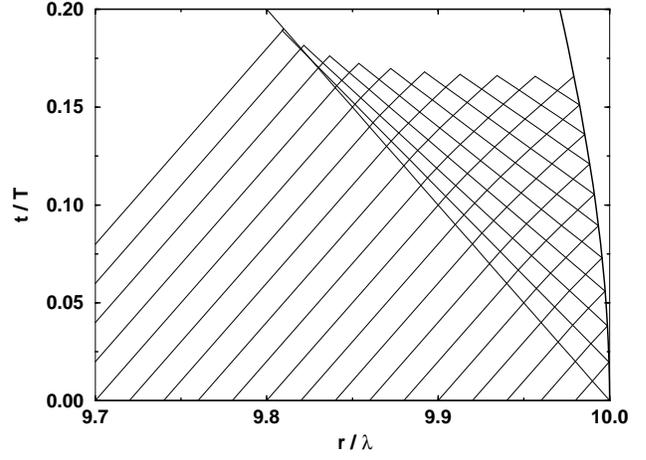}}
\caption{
Characteristics obtained
for an experimental oscillation amplitude equal to
$\Delta x_0 = 10$ $\mu$ m, a cell radius $\RL_0 / \lambda = 10$,
and a time step $\delta t = 0.01$ T.
The location of the piston is represented by the solid thick
line on the right. On the upper left, two characteristics
are crossing each other and the program stops.
We chose a much greater oscillation than in the
experiment, in order to have  a rapid shock formation,
and thus a readable picture.
}
\label{fig_car4}
\end{figure}

We shall now show that such an intersection does occur
in the system, and calculate the corresponding shock length.
We consider a spherical domain bounded by a spherical piston.
At $t=0$, the fluid is at rest with a density $\RHOST$
(label ``st'' standing for static) and the piston
surface is a sphere of radius $\RL_0$.
As long as the fluid is at rest, all the characteristics
are straight lines
(see Fig.~\ref{fig_car4}).
  The respective slopes of the characteristics $\left\{\CCUN_i\right\}_{i \in R}$ and $\left\{\CCD_i\right\}_{i \in R}$
are $+c_\mathrm{st}$ and $-c_\mathrm{st}$,
$c_\mathrm{st}$ being the sound velocity for the initial density
$\RHOST$.
The piston starts to move at $t=0$ with a velocity
$v_p(t) = -\Delta v_0 \sin (\omega t)$.

We denote $\CCD_0$ the characteristic
which originates from $r=\RL_0$ when t=0, with slope $-c_\mathrm{st}$.
The domain to the left of $\CCD_0$ is unperturbed
unless some characteristic $\CCD_i$
crosses $\CCD_0$, i.e. unless there is a shock.

The aim of this calculation is to find an upper bound
for the time necessary to form a
first shock in the system.
As the piston moves, it emits some characteristics
which will cut $\CCD_0$ after a while, leading to a shock.
We only study how the characteristics emitted at early times,
and almost parallel to $\CCD_0$, will eventually cross it.
Of course, some other characteristics emitted later could
cross $\CCD_0$ earlier, or some shocks could occur somewhere else
at earlier times. That is why our calculation
only gives an upper bound for the shock time.

The details of the calculation are given in Appendix~B.
We find that an upper bound for the time of shock
formation is
\begin{equation}
t_\mathrm{shock} \le \frac{\RL_0}{c_\mathrm{st}} \left\{ 1 -
\exp\left[ -\frac{c_\mathrm{st}^2}{2\; \RL_0\; \omega\; \Delta v_0}
\;\frac{\RHOST-1}{\RHOST-\frac{1}{2}} \right] \right\}
\le \frac{\RL_0}{c_\mathrm{st}}.
\end{equation}
As the corresponding shock length $r_\mathrm{shock}$
is measured from the center of the
sphere, a lower bound for $r_\mathrm{shock}$ is
\begin{equation}
\RL_0 \ge r_\mathrm{shock} \ge \RL_0 \exp\left[
  -\frac{c_\mathrm{st}^2}{2\; \RL_0\; \omega\; \Delta v_0}
\;\frac{\RHOST-1}{\RHOST-\frac{1}{2}} \right] > 0.
\label{rchoc}
\end{equation}

\begin{figure}
\centerline{\includegraphics[width=8.0cm]{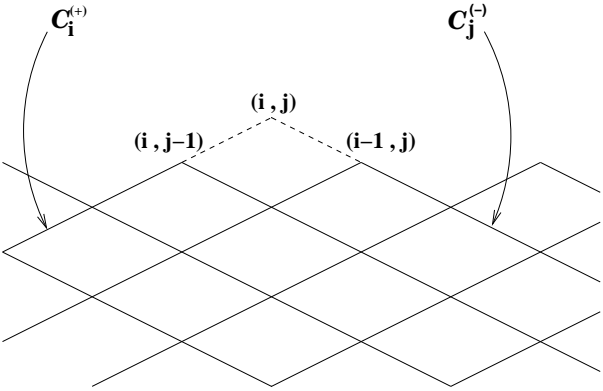}}
\caption{\em Schematic representation of the network
formed by the two families of characteristics, where
$i$ and $j$ are integers.
}
\label{fig_scheme}
\end{figure}

From this study we conclude that in spherical geometry,
there is {\em always} formation of a shock, whatever the
velocity of the piston is, as long as it has a nonzero
acceleration towards the
center of the sphere.
However, when the oscillation amplitude $\Delta x_0
= \Delta v_0 / \omega$ goes to zero, the shock forms
very near the focal point.
As shocks are formed by the intersection of tangential
characteristics, their initial amplitude is zero.
If they are formed very near the center, their amplitude
does not have time to grow much.
This is true especially because of the existence of a cut-off:
the notion of shock becomes meaningless when the width
of the shock becomes of the same size as the shock length
itself.
So, as the oscillation amplitude tends to zero,
the jump in density and velocity at the shock also
vanishes, and their is no contradiction with the
fact that the solution is expected to approach
the linear solution~\cite{landau_l6}.

The above calculation is analytic and gives
only a lower bound for the shock length
because we have assumed that the shock occurs
on the first characteristic $\CCD_0$.

\subsubsection{Numerical calculation of the shock length}

As we shall see now, a numerical simulation shows that our calculation of
the lower bound is in fact a good estimate of the shock length itself.
Let us now solve Euler's equations numerically by computing
a network of characteristics $\left\{\CCD_i\right\}_{i \in Z}$
and $\left\{\CCUN_i\right\}_{i \in Z}$.
The values of the Rieman invariants, and thus of the density
$\rho$ and velocity $u$, will be defined on the intersections
of the left $\CCD_j$ and right $\CCUN_i$ characteristics.
As the fluid is at rest on the left of $\CCD_0$, we
restrict our calculation to the $(r,t)$ domain located
between $\CCD_0$ and the piston trajectory.

Several parameterization choices are possible.
For the $(+)$ family, we have chosen to take $i = t_i/\delta t$
for each characteristic $\CCUN_i$, where $t_i$ is the
time at which $\CCUN_i$ and $\CCD_0$ intersect,
and $\delta t$ is an arbitrary fixed time step.
We discretize the system by restricting $i$ to
integer values.
This means that the subset of characteristics that will
be considered for the simulation are initialized at regular
time intervals $i \delta t$ when they cross the first
characteristic $\CCD_0$ emitted by the piston
(see Fig.~\ref{fig_car4}).

When $\CCUN_i$ meets the piston, a new characteristic
$\CCD_i$ (with the same label $i$) is emitted
from the piston at the same time.
This defines the parameterization of the $(-)$ family.
The intersections of these two families of characteristics
form a network of points whose locations will be determined
in the following.
The intersection of the characteristics $\CCUN_i$ and
$\CCD_j$ is referred to as point $(i,j)$.

During the $n^{th}$ step, we compute all the points
$(i,j)$ such that $i+j=n$ with $i$ and $j$ integers.
Let us describe now how a point $(i,j)$ can be
computed from the points of the previous step
(see Fig.~\ref{fig_scheme}).
All the information comes from the two sites
$(i,j-1)$ and $(i-1,j)$.
We must extrapolate each characteristic $\CCUN_{i}$
and $\CCD_{j}$ up to the next intersection $(i,j)$.
We compute the local slopes $u \pm c_\mathrm{s}$ of the characteristics
$\CCUN_{i}$
and $\CCD_{j}$ in sites $(i,j-1)$ and $(i-1,j)$
respectively.
Then $(i,j)$ is the intersection of the two straight lines
which respectively go through $(i,j-1)$ and $(i-1,j)$
with these slopes.

The values of the Rieman invariants $\INVUN$ and $\INVD$
at $(i,j)$ are found by numerical integration of
the equations (\ref{defIk},\ref{eq_char}) along the two involved
characteristics.
 From these values, both $\rho$ and $j=\rho u$ can be obtained.

The program stops whenever two characteristics
of the same family cross
each other, as shown on Fig.~\ref{fig_car4}.
Then a shock occurs and the calculation
based on characteristics breaks down (there would be
multivalued points).

\begin{figure}
\centerline{\includegraphics[width=8.5cm]{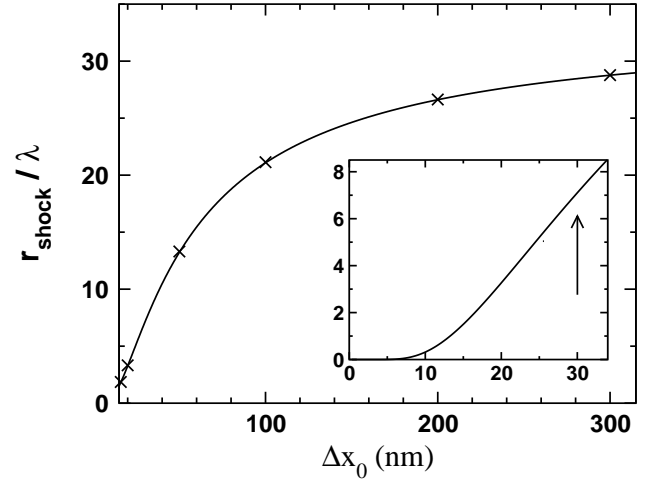}}
\caption{\em Shock distance in helium~4 versus the amplitude
of the oscillation on the ceramic,
for a cell radius $\RL_0 = 33.6 \lambda$.
We compare the analytic prediction
(solid line) and our numerical calculations
(symbols),
both based on the characteristics method. The inset shows an
expanded view for small amplitudes.
The arrow indicates the amplitude for
which cavitation is observed in the experiment, starting from
a zero static pressure.}
\label{fig_char}
\end{figure}

Let us now compare analytic and numerical results.
Both calculations were done by taking $\RL_0$
equal to the experimental cell radius,
i.e. 8 mm = 33.6 $\lambda$.
In Fig.~\ref{fig_char}, we plot the shock distance $r_\mathrm{shock}$
measured from the center of the sphere
as a function of the amplitude of the oscillation of the
piston. We compare it with the analytic result.
The agreement is excellent.
This shows that our analytic calculation not only gives a lower bound
but in fact a good estimate for the shock distance itself.

We have also performed the analytic calculation in
the case of helium~3, neglecting the role of viscosity.
It is expected that shocks form at smaller wave amplitude in helium 3,
because, in this lighter liquid, the spinodal pressure is less negative
than in helium 4.

\subsection{Numerical simulations of Euler's equation}

\label{sect_simul}

\subsubsection{Numerical method}

\label{sect_finite}

\begin{figure}
\centerline{\includegraphics[width=8.5cm]{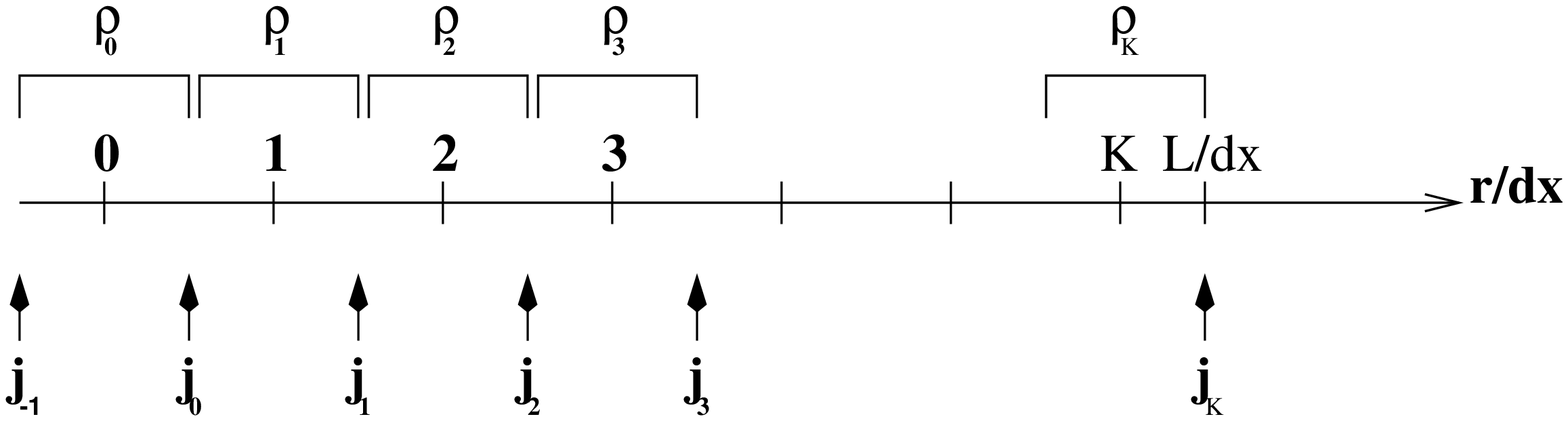}}
\caption{\em Staggered lattice used for the numerical simulations.
The radius of the simulation domain
is $\RL=\left(K+\frac{1}{2}\right) \delta r$.
The symmetry with respect to the center $r=0$ imposes
$j_0 = -j_{-1}$, so that the velocity would vanish
at the center : j(r=0) = 0.}
\label{fig_mail}
\end{figure}

As we shall see now, a simple finite difference numerical scheme
is sufficient to simulate our system with moderate
amplitude, at least as long as the shock length
is smaller than the spatial discretisation step.
Our aim is to calculate
numerically the pressure and density oscillation at the center.
We chose to have two lattices,
one for mass and the other for momentum.
  They are staggered (Fig.~\ref{fig_mail}) and allow us to enforce exactly
the conservation of mass:

\begin{equation}
\rho^{t+\delta t/2}_k = \rho^{t-\delta t/2}_k + \frac{\delta t}{\delta r}
\frac{\left[ \left(k-\frac{1}{2}\right)^2 j^t_{k-1}
- \left(k+\frac{1}{2}\right)^2 j^t_k \right]}
{k^2 + \frac{1}{12}}.
\end{equation}
We have taken the momentum equation in the form:
\begin{eqnarray}
j^{t+\delta t}_k & = & j^t_k - \frac{\delta t}{\delta r}
\left[ \frac{ (j^t_k + j^t_{k+1})^2}{4 \rho^{t+\delta t/2}_{k+1}}
- \frac{ (j^t_k + j^t_{k-1})^2}{4 \rho^{t+\delta t/2}_k} \right]
\nonumber \\
& - & C_0^2
\left( \frac{\rho^{t+\delta t/2}_k + \rho^{t+\delta t/2}_{k+1}}{2}
- 1 \right)^2 \frac{\rho^{t+\delta t/2}_{k+1} - \rho^{t+\delta t/2}_k}{\delta r}\delta t
\nonumber \\
& - & 4 \frac{\delta t}{\delta r} \frac{(j^t_k)^2 }{\left(k+\frac{1}{2}\right)
\left[ \rho^{t+\delta t/2}_k + \rho^{t+\delta t/2}_{k+1} \right]}.
\end{eqnarray}
As we use a staggered lattice, we only have to specify
the boundary conditions for the momentum.
It is vanishing at the center of the sphere (r=0),
and thus the symmetry with respect to the center
imposes $j_0 = -j_{-1}$ (see Fig.~\ref{fig_mail}).
On the other hand, the motion of the piston is implemented
by
$$
j_K(t) = - \rho_K \, \omega \, \Delta x_0
\, \sin (\omega t)
\, \left[1 - \exp(-t/1.5)\right].
$$

\subsubsection{Focal pressure}

\label{sect_focal}

\begin{figure}
\centerline{\includegraphics[width=8.0cm]{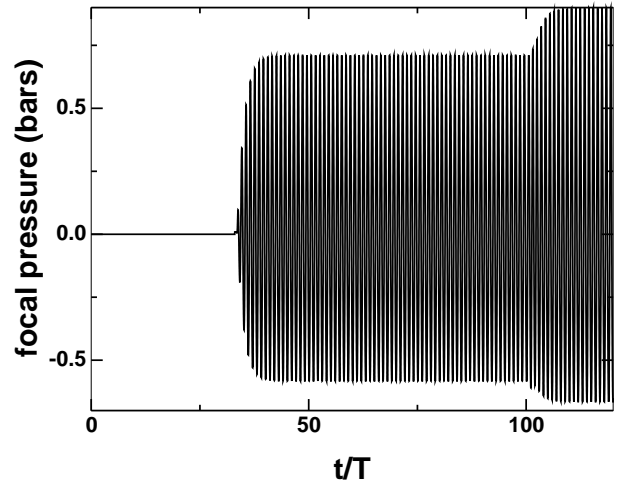}}
\caption{\em Focal pressure for a weak oscillation $\DELTAX = 0.7$ nm.
The simulation was done with $100$
mesh points per wavelength,
starting with a static density $\RHOST = \rho_0$,
and thus a vanishing static pressure.
}
\label{fig_pc1}
\end{figure}

In Fig.~\ref{fig_pc1}, the focal pressure
is represented as a function of time.
It is calculated from the average density
in the central cell.
The results of Fig.~\ref{fig_pc1} were
obtained for a cell length equal to
the experimental one, i.e. $\RL_0 = 33.6\; \lambda$,
an oscillation amplitude $\Delta x_0 = 0.7$ nm,
and a zero static pressure.
Then the reduced sound velocity is equal to 1,
and the wave needs 33.6 time units to reach the
center of the cell.
 We are interested in the steady
regime, which is established around $t/T > 45$.
For $t/T > 33.6$, the wave reflected on the
focal point propagates
towards the ceramic, and then back to the focal
point. It reaches the latter at $t/T = 100.8 = 3 \times 33.6$.
Then we stop our measurements, to be consistent
with the experiment, which uses short bursts.

\begin{figure}
\centerline{\includegraphics[width=8.0cm]{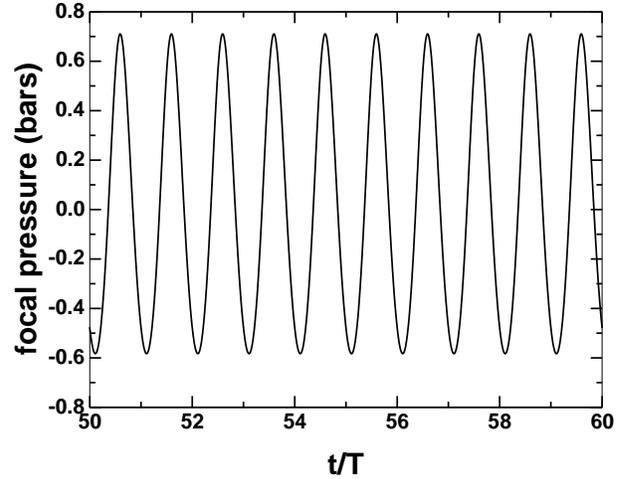}}
\caption{\em Same figure as Fig.~\protect{\ref{fig_pc1}},
with a zoom on the steady state region.
We are nearly in the linear regime.
}
\label{figlinearpres}
\end{figure}

As long as the oscillation amplitude of the ceramic
is not too large, non-linear effects are negligible
far from the focal point.
If we observe the focal pressure during the steady
regime, we find it also nearly sinusoidal,
and the positive swings are only slightly larger
than the negative ones
(Fig.~\ref{figlinearpres}).

\begin{figure}
\centerline{\includegraphics[width=8.5cm]{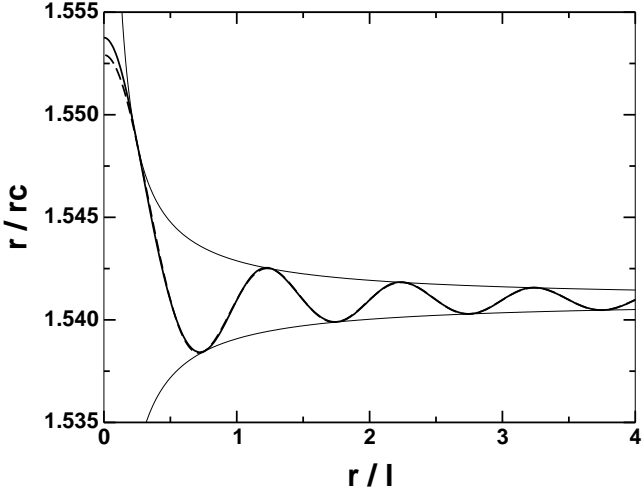}}
\caption{\em Density profile for a weak oscillation
$\DELTAX = 0.7$ nm
(same simulation as in Fig.~\protect{\ref{fig_pc1}}).
We are still in the linear regime. The thin solid lines indicate
a fit of the amplitude by $1/r$ and the dashed line a fit
by $\sin(kr)/(kr)$. The latter is almost identical to the
numerical result, except near the focal point where non-linear
effects become visible. The simulation was done with $100$
mesh points per wavelength, starting with a static density
$\RHOST = \rho_0$ - and thus a vanishing static pressure.
}
\label{figlinearprof}
\end{figure}

We have checked that the amplitude of the density
is well represented by the function $\sin(kr)/(kr)$, as predicted
from the linear
theory~\cite{landau_l6} (see Fig.~\ref{figlinearprof}).

 From our calculation of section \ref{sect_charact},
we expect a shock to occur near the focal point.
However, for the case of Fig.~\ref{figlinearprof},
it happens on a region around $r=0$ that cannot
be seen because it is much smaller than the mesh size.
Indeed, the predicted reduced shock length would be
$r_c / \lambda = 4.2 \times 10^{-28}$, to be compared
with the mesh size $\delta r / \lambda = 0.01$.

\begin{figure}
\centerline{\includegraphics[width=8.5cm]{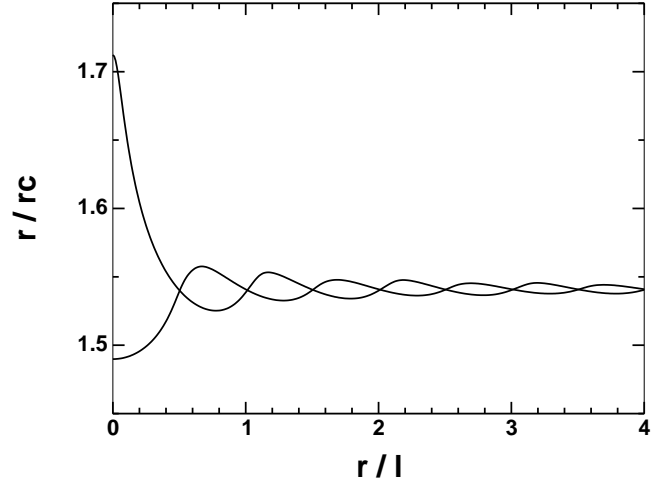}}
\caption{\em Density profile for an oscillation $\DELTAX = 5$ nm,
at different times, corresponding to the maximal and
minimal focal pressure.
Non linearities are becoming important.
The simulation was done with $100$
mesh points per wavelength,
starting with a static density $\RHOST = \rho_0$.
}
\label{fignonlinearprof0}
\end{figure}

\begin{figure}
\centerline{\includegraphics[width=8.0cm]{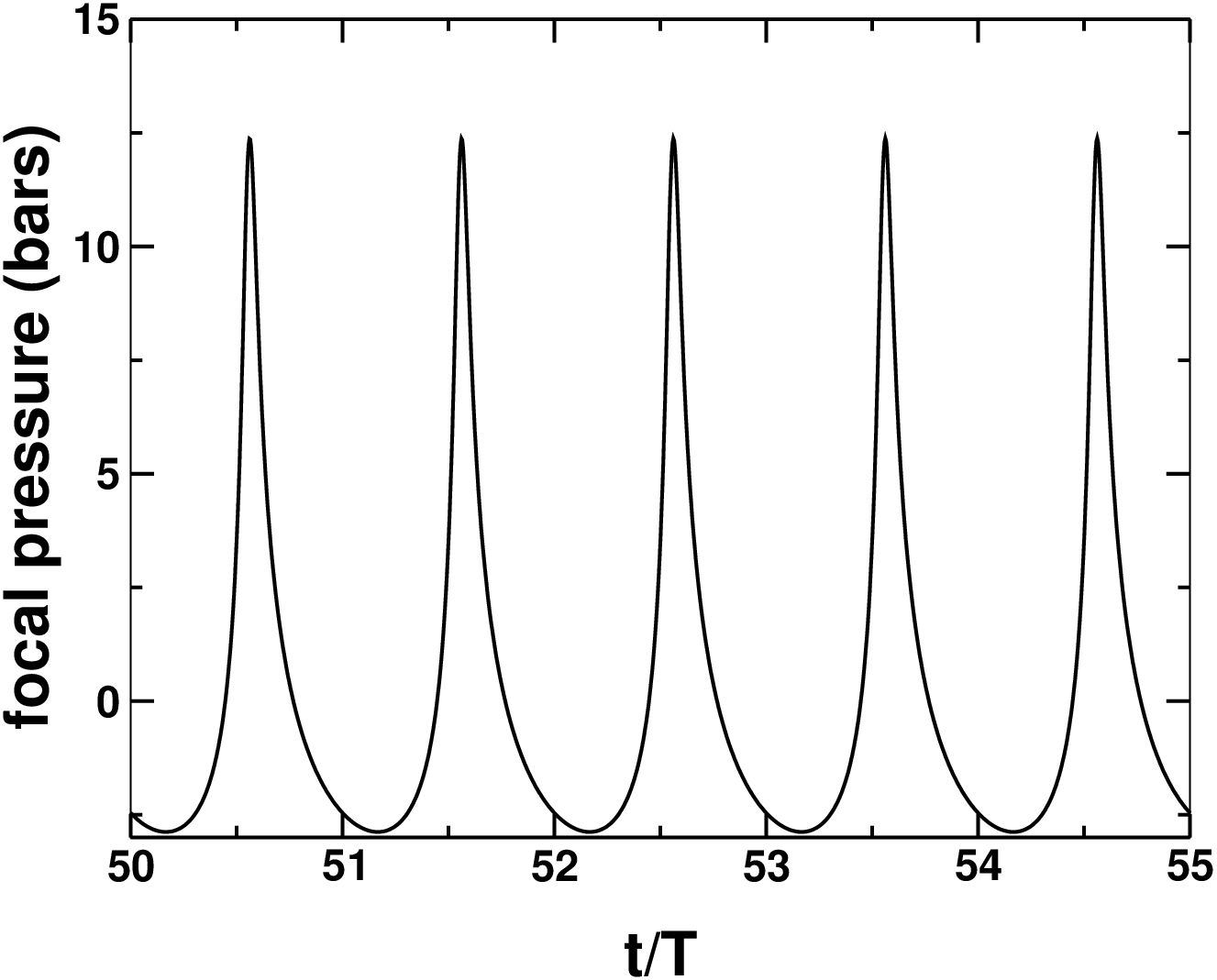}}
\caption{\em Focal pressure for an oscillation $\DELTAX = 5$ nm.
Non linearities are becoming important.
The simulation was done with $100$
mesh points per wavelength,
starting with a static density $\RHOST = \rho_0$.
}
\label{fignonlinearpres0}
\end{figure}

When $\DELTAX$ is increased to $5$ nm, non-linear effects
become more important.
One sees the formation of fronts
(Fig.~\ref{fignonlinearprof0}).
The reduced shock length (from the center)
is equal to $0.003$ and is thus still smaller
 than the spatial step $\delta r/\lambda = 10^{-2}$.
At the center, the positive swings of the pressure
are now much larger than the negative ones
(Fig.~\ref{fignonlinearpres0}).
There are several sources for non linearities:
the equation of state; the inertial term in Euler
equations; both enhanced by the spherical geometry.
The relative importance of these factors will be
discussed now.
We have performed various simulations where we suppressed one term or
the other.
Results are summarized in Table~\ref{tab_nonlin}.

If we take a constant sound speed
$c_\mathrm{s} = c_\mathrm{st}$, i.e. a linear equation of state,
then non-linear effects are strongly reduced and the maximal
and minimal pressure excursions are almost symmetrical.

If we rather suppress the inertial term
$u\cdot\nabla u$ from the Euler equation for the velocity,
non-linear effects are also reduced, but to a lesser extent.

\begin{table}
\centerline{
\begin{tabular}{cccccc}
\\
\hline
$\RL_0$ / $\lambda_0$ & $\RL_0$ (mm) & & case 1 & case 2 & case 3 \\
10   & 2.38 & $P_{max}$ & 13.53 & 6.18 & 11.92 \\
& & $P_\mathrm{min}$ & -3.49 & -4.83 & -3.58 \\
20   & 4.76 & $P_{max}$ & 17.17 & 6.34 & 14.27 \\
& & $P_\mathrm{min}$ & -3.31 & -4.73 & -3.41 \\
\hline
\end{tabular}
}
\caption{Maxima and minima of the focal
pressure (bars) computed in three different cases.
Case 1: full simulation of the equations (\ref{eqeulerj}).
Case 2: the sound speed is kept constant $c_\mathrm{s} = c_\mathrm{st}$.
Case 3: the inertial term $u\cdot\nabla u$ is suppressed.
All calculations are done for the same experimental
oscillation amplitude $\Delta x_0 = 5.9\,\mathrm{nm}$, and for
two different cell radii.}
\label{tab_nonlin}
\end{table}

Thus all non-linear terms reinforce each other, but
the dominant non-linear effect comes from the
equation of state (the geometry also plays a crucial
role of course, which cannot be separated from the other
effects).

\begin{figure}
\centerline{\includegraphics[width=8.5cm]{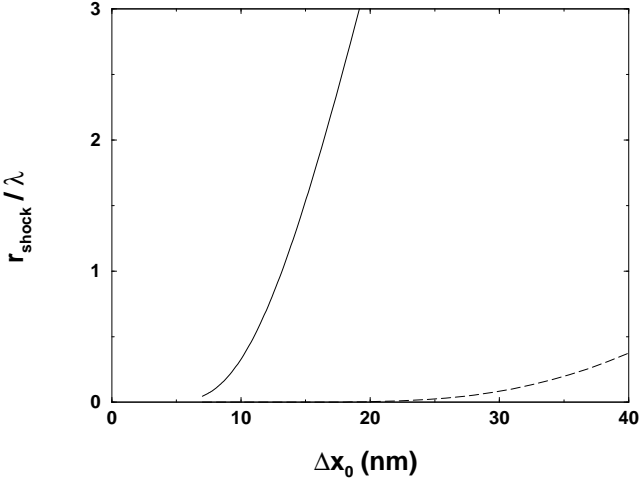}}
\caption{\em Shock distance in helium~4 versus the amplitude
of the oscillation of the ceramic.
We compare the analytic prediction
for the full equation of state
(solid line) and the case of a constant sound speed
(dashed line).
}
\label{fig_nonlin}
\end{figure}

This is confirmed if we do the analytic calculation of
section \ref{sect_analytic} again, now with a constant sound speed.
We find then that
\begin{equation}
r_\mathrm{shock} \simeq \RL_0 \exp\left[ -\frac{c_\mathrm{st}^2}{\RL_0 \omega \Delta v_0}
\right].
\label{rchoc_cs0}
\end{equation}
On Fig.~\ref{fig_nonlin}, we compare with the result
(\ref{rchoc}) of section
\ref{sect_analytic} obtained with the full equation of state.
It turns out that, for realistic oscillation amplitudes,
the shock appears almost at the center if the nonlinearities
of the equation of state are not taken into account.

\subsection{Shock formation at large amplitudes. The WENO scheme}
\label{sect_weno}

Eventually, when the amplitude $\DELTAX$ is further increased,
the formation of shocks can be observed in our simulations.
The numerical scheme described in the previous section
becomes unstable when $r_\mathrm{shock} \simeq \delta r$, i.e.
for a vanishing static pressure
and a reduced $\delta r = 10^{-2}$,
when $\DELTAX \simeq 6\,\mathrm{nm}$.
Then a new numerical scheme has to be used.

For performing fine analysis of the Euler flow dynamics, the
numerical scheme must recover low dissipative pro\-perty.
Non-dissipative high-order accurate schemes (like spectral or
Pad\'e schemes) have been identified as suitable tools as far as
regular numerical solutions are searched. Nevertheless, when
dealing with compressible flows involving discontinuities,
non-dissipative high-order schemes introduce spurious oscillations
in the vicinity of the discontinuity and one must use a numerical
scheme which can both represent the smooth regions of the solution
with the minimum of numerical dissipation, and capture the
discontinuities by using an {\em ad hoc} scheme with a robust
discontinuity-capturing features. Therefore, as shock waves may
occur in the computational domain for the present calculations,
the numerical method we use is based on a high-order
intrinsically-dissipative scheme originally designed to capture
discontinuities. The method, called WENO (Weighted Essentially
Non-Oscillatory)~\cite{jiang_s96}, is presented in the next
section, and some more technical details about WENO and ENO
schemes can be found in the literature~\cite{jiang_s96,tenaud_g_s00,shu_o89,shu_o88,roe81}

As in many schemes devoted to computations involving shocks, the
WENO scheme uses the Riemann invariants as variables. Computing
the evolution of these variables requires to  know their values
not only at integer space coordinates, but also at some
intermediate locations. An extrapolation from integer positions is
thus necessary and this is where the fundamental idea of WENO
schemes comes in. The time evolution scheme (a third order
Runge-Kutta) and the change of variables towards the
characteristic plane are more classical, but we shall still recall
them for self-consistency.

\subsubsection{The WENO method in more details}
\label{subsect_simul}

For simplicity, the governing equations (\ref{eqeulerj})
are recast in the following abridged form:
\begin{equation} \label{euler_vector}
\frac{\partial Q}{\partial t} = {\cal L} \left( r,Q \right)
\end{equation}
where $\displaystyle{Q = \left( \rho \ , \ \rho \ u \right) ^t }$
is the vector of the conservative variables and
$$\displaystyle{ {\cal L} \left( r, Q \right) = - \frac{\partial F
\left( Q \right)}{\partial r} + {\cal S} \left( r, Q \right) }$$
stands for a spatial operator, applied on $Q$, based on both the
Euler flux vector $\displaystyle{ F \left( Q \right) = \left( \rho
\ u \ , \ \rho \ u^2 + P \right) ^t}$ and the source term
$\displaystyle{ {\cal S} \left( r, Q \right) = - \frac{\theta}{r}
\left( \rho \ u  , \rho \ u^2 \right) ^t }$.

In view of the discretization of the Euler equations
(\ref{euler_vector}), we will denote by $\delta t$ and $\delta r$
the time step and cell width respectively. $Q_{i}^{n}$ will denote
the numerical vector solution at a time $t = t_{0} + n \cdot
\delta t$ and at a position $r = i \cdot \delta r$. For
simplicity, the integration of these equations has been performed
by means of a decoupled time and space algorithm.

\vskip 0.2cm
\noindent
{\bf a) Time integration}
\vskip 0.2cm

The time integration is then performed by means of a third-order
accurate Runge-Kutta method, proposed by Shu and
Osher~\cite{shu_o89}, chosen because this high order accurate
scheme does not increase the Total Variation of the
right-hand-side of the equations (${\cal L} \left( r,Q \right)$).
When dealing with discontinuities, this Total Variation
Diminishing (TVD) property is important since it ensures that no
local extremum can be created during the time integration, meaning
that non oscillation may occur in the shock wave vicinity due to
the time scheme. At each point of the computational grid, the time
integration is then obtained via a multi-step algorithm as
follows:
\begin{eqnarray}
\nonumber Q^{(0)} & = & Q_i^n \\
\nonumber Q^{(1)} & = & Q^{(0)} + \delta t \ {\cal L} \left( Q^{(0)} \right) \\
          Q^{(2)} & = & \frac{3}{4} \ Q^{(0)} + \frac{1}{4} \ Q^{(1)}
       +  \frac{1}{4} \delta t \ {\cal L} \left( Q^{(1)} \right) \\
\nonumber Q^{(3)} & = & \frac{1}{3} \ Q^{(0)} + \frac{2}{3} \ Q^{(2)}
+  \frac{2}{3} \delta t \ {\cal L} \left( Q^{(1)} \right) \\
\nonumber Q_i^{n+1} & = & Q^{(3)}
\end{eqnarray}

The explicit Runge-Kutta scheme exhibits a stability condition
based on the $CFL$ number: $CFL = \mu \ \delta t / \delta r$,
where $\mu$ is the maximal value of the eigenvalues $\mu^k_i$,
to be defined below. All
the calculations presented herein have been obtained by
considering $CFL=0.5$, which corresponds to a nearly optimal value
for the considered Runge-Kutta scheme. Let us mention that this
value ensures a good representation of all the time scales of the
flow considered.

\vskip 0.2cm
\noindent
{\bf b) Spatial integration}
\vskip 0.2cm

The spatial discretization of the right-hand-side term ${\cal L}
\left( r,Q \right)$ of equations~(\ref{euler_vector}) is obtained
by means of a high-order finite difference scheme:
\begin{equation} \label{euler_discrete}
{\cal L} \left( r_i , Q_i^n \right) = - \frac{1}{\delta r} \left[
\overline{F}^n_{i+1/2} - \overline{F}^n_{i-1/2} \right] + {\cal S}
\left( r_i , Q_i^n \right)
\end{equation}
where $\displaystyle{\overline{F}^n_{i+1/2} }$ is the numerical
flux evaluated at the cell interface ($r_{i+1/2}$). To reconstruct
the numerical flux at the cell interfaces, a scheme with a
discontinuity-capturing feature must be employed  to prevent
oscillation in the vicinity of the shock wave. Following a
previous study~\cite{tenaud_g_s00} on the capability of some recent
high-order shock capturing schemes to recover basic fluid mechanic
phenomena, the numerical flux has been evaluated by means of
Essentially Non-Oscillatory (ENO) family scheme~\cite{jiang_s96,shu_o89,shu_o88}. The numerical flux
is approximated by means
of polynomial reconstruction over several grid points
(the set of these points is named
``stencil") around the cell interface considered.
We shall now describe this reconstruction of the
fluxes in details.

$\bullet$ {\bf Change of variable:} For simplicity and accuracy
purposes, the discretization of the Euler flux is based on a
polynomial reconstruction applied on the local characteristic
variables (Riemann invariants, see \S~\ref{sec_rieman}) since the
equations recover the scalar form. Then, the propagation
directions can easily be followed in the characteristic plane. In
order to perform this change of variables, the method is the same
as in Appendix~A. First, one linearizes the Euler equations and
finds the eigenvectors of the Euler flux jacobian evaluated at the
cell interface ($\displaystyle{\left(
\partial F / \partial Q \right) _{i+1/2} }$).
Note that, as we are now using the Euler equations
in conservative form (\ref{eqeulerj}) instead of
(\ref{eqeuler}), the jacobian differs from
the one given in appendix A (matrix $A$), and
reads now
\begin{equation}
\partial F / \partial Q =
\left( \begin{array}{ccc}
\displaystyle{0} &\;\;& \displaystyle{1} \\
\displaystyle{{-j^2\over \rho^2}+c_\mathrm{s}^2} &\;\;& \displaystyle{2 {j\over \rho}}
\end{array}
\right)
\end{equation}
In order to compute
$\displaystyle{\left(
\partial F / \partial Q \right) _{i+1/2} }$.
- and then the eigenvalues
($\mu^{k}_{i+1/2})_{k=\pm}$, and the left ($\mathbf{l}^{k}_{i+1/2}$) and
right ($\mathbf{r}^{k}_{i+1/2}$) eigenvectors -,
the conservative variables $Q^n_{i}$
must be evaluated at the cell interface $r_{i+1/2}$
(note that a bold $\mathbf{r}^{k}_{i+1/2}$ refers to the
eigenvector, while $r_{i+1/2}$ stands for the scalar
spatial coordinate).
As these variables do not vary linearly in the cell
if a shock is present, one cannot use a simple
arithmetic or geometric average, but rather
a Roe average, whose description can be found in Ref.~\cite{roe81}.
This ensures the consistency of the scheme, i.e. that
$\displaystyle{\left(
\partial F / \partial Q \right) _{i+1/2}} $
converges towards
$\displaystyle{\left(
\partial F / \partial Q \right) _{i} }$
when $\delta r$ tends to zero.

The numerical Euler flux is then
projected onto the left eigenvector matrix
($\displaystyle{{\bf l}^{(+)}_{i+1/2}, {\bf
l}^{(-)}_{i+1/2} }$).
The scalar ENO
reconstruction procedure is applied to the projected
fluxes~\cite{shu_o89}. In the physical domain, the numerical
Euler flux are then obtained by a projection onto the right
eigenvectors \linebreak
($\displaystyle{ {\bf r}_{i+1/2} = {\bf
l}^{-1}_{i+1/2} }$) and read:
\begin{equation} \label{ENO1}
\overline{F}_{i+1/2}  =  \sum_{k=\pm} \left[ {f^{ENO}}^k_{i+1/2}
\cdot \mathbf{r}^k_{i+1/2} \right]
\end{equation}
where ${f^{ENO}}^k_{i+1/2}$ stands for the scalar ENO
reconstruction, which will be defined below.

$\bullet$ {\bf Extrapolation of variables at non-integer
locations - the core of (W)ENO schemes:}

\begin{figure}
\centerline{\includegraphics[width=8.5cm]{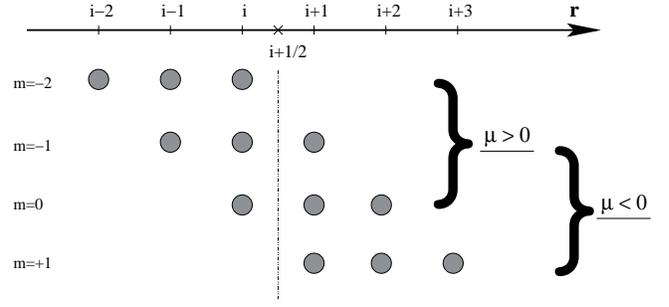}}
\caption{\em Sketch of all the stencil candidates to recover a
third-order reconstruction ($p = 3$) at the cell interface
$r_{i+1/2}$. The eigenvalue $\mu$ stands for the
extrapolated value $\mu^k_{i+1/2}$.} \label{fig_stencil}
\end{figure}

First, we shall present the strategy used in ENO (Essentially
Non--Oscillatory) schemes~\cite{shu_o89}. For a given non integer
location, there are several ways to perform the extrapolation,
depending on from how many integer points it will be performed,
and how these points will be located with respect to the
non-integer one. The set of these points is called a stencil.
Fig.~\ref{fig_stencil} illustrates the different possible
choices for a stencil of length $p=3$. Simple finite difference
schemes use a stencil defined once for all. Here, only the length
$p$ of the stencil is fixed for a given simulation. All the $p+1$
possible locations are considered as candidates, provided that
there is at least one point adjacent to the non-integer point (see
again Fig.~\ref{fig_stencil}).

If we denote by $p$ the order of the reconstruction, the ENO
procedure~\cite{shu_o89} chooses the most regular stencil among
the $p+1$ stencil candidates. As an exemple for $p = 3$, we can
see all the stencil candidates on Fig.~\ref{fig_stencil}.
A first selection among the $p+1$ stencil candidates
is performed according to
the sign of the two eigenvalues ($\mu_{i+1/2}^{k}$):
one keeps the $p$ most left stencils for the
positive or null eigenvalues and the $p$ most right otherwise
(Fig.~\ref{fig_stencil}). Indeed, the sign of the eigenvalue
gives the propagation direction of the associated
characteristics, and thus of the relevant information.

The regularity of the function on
each of the $p$ remaining stencils
is measured
by the undivided difference Table~\cite{shu_o89}
evaluated on each stencil and the
most regular stencil is chosen among all the $p$ stencil
candidates.
Of course,
this stencil may be different at each time step,
for each location, and for each eigenvalue
$(\mu_{i+1/2}^{k})_{k=\pm}$.

The scalar ENO reconstruction is then applied on this
specific stencil by means of the following polynomial development:
\begin{equation} \label{ENO2}
{f^{ENO}}^{k,m}_{i+1/2} =  \sum_{j=0}^{p-1} \zeta^{p,k}_j
\mathbf{l}^{k}_{i+1/2} \cdot F \left( Q_{i+m+j} \right)
\end{equation}
The integer $m$ refers to the index of the most left point of the
chosen stencil.
The sum goes over all the points of this stencil.
One recognizes the projection of the fluxes onto the left eigenvectors,
that allows to go from physical variables to characteristic
variables. The constant coefficients of the polynomial
($\zeta^{p,k}_j $) are calculated in order to recover
a scheme of order $p$ in regular regions.
The values of $\zeta^p_j$ can be found in Table~\ref{table_zeno}
up to $p = 5$.

\begin{table}
\centerline{
\begin{tabular}{||c|c|c|c|c|c|c||}
\hline \hline
$p$ & $m$ & $j=0$ & $j=1$ & $j=2$ & $j=3$ & $j=4$ \\
\hline \hline
1 & -1 & 1 & & & & \\
\hline  & 0 & 1 & & & & \\
\hline \hline 2 & -1 & -1/2 & 3/2 & & & \\
\hline  & 0 & 1/2 & 1/2 & & & \\
\hline  & +1 & 3/2 & -1/2 & & & \\
\hline \hline 3 & -2 & 1/3 & -7/6 & 11/6 & & \\
\hline  & -1 & -1/6 & 5/6 & 1/3 & & \\
\hline  & 0 & 1/3 & 5/6 & -1/6 & & \\
\hline  & +1 & 11/6 & -7/6 & 1/3 & & \\
\hline \hline 4 & -3 & -1/4 & 13/12 & -23/12 & 25/12 & \\
\hline  & -2 & 1/12 & -5/12 & 13/12 & 1/4 & \\
\hline  & -1 & -1/12 & 7/12 & 7/12 & -1/12 & \\
\hline  & 0 & 1/4 & 13/12 & -5/12 & 1/12 & \\
\hline  & +1 & 25/12 & -23/12 & 13/12 & -1/4 & \\
\hline \hline 5 & -4 & 1/5 & -21/20 & 137/60 & -163/60 & 137/60 \\
\hline  & -3 & -1/20 & 17/60 & -43/60 & 77/60 & 1/5 \\
\hline  & -2 & 1/30 & -13/60 & 47/60 & 9/20 & -1/20 \\
\hline  & -1 & -1/20 & 9/20 & 47/60 & -13/60 & 1/30 \\
\hline  & 0 & 1/5 & 77/60 & -43/60 & 17/60 & -1/20 \\
\hline  & +1 & 137/60 & -163/60 & 137/60 & -21/20 & 1/5 \\
\hline \hline
\end{tabular} }
\caption{The constant coefficients $\zeta^p_j $ of the ENO
reconstruction up to 5$^{th}$-order.} \label{table_zeno}
\end{table}

Using this generic ENO scheme~(\ref{ENO1}, \ref{ENO2}), the Euler
flux derivative is estimated with a $p^{th}$-order of accuracy at
best (in regular regions).
However, when the stencil used at the cell interface
$r_{i+1/2}$ is different from the one at $r_{i-1/2}$
(which is the case in strong gradients or shock regions),
the order of accuracy decreases.

One of the drawbacks in the generic ENO scheme is the necessity to
check and choose between $p$ stencil candidate, which
is quite CPU consuming.
To overcome this
disadvantage, we preferred using a Weighted ENO scheme (WENO)
since it improves the order of accuracy of the generic ENO scheme
by using a weighted combination of the $p$ possible stencils. The
weights~\cite{jiang_s96} depend on the degree of regularity of
the solution. In regular regions, they can be computed to achieve
$(2p-1)^{th}$-order of accuracy whereas in regions with
discontinuities they are set to zero, leading to a standard ENO
scheme. The WENO flux is estimated by:
\begin{equation} \label{WENO1}
{f^{WENO}}^k_{i+1/2} =  \sum_{n=0}^{p-1} \omega _n  \
{f^{ENO}}^{k,n-p+1}_{i+1/2}
\end{equation}
where ${f^{ENO}}^{k,n-p+1}_{i+1/2}$ is the generic ENO
reconstruction given by equation (\ref{ENO2}) and $\omega _n$ are
the weights defined as follows:
\begin{eqnarray} \label{WENO2}
& \displaystyle{ \sum^{p-1}_{n=0} \omega _n =  1,} & \\
\omega _n  = \frac{\beta_n}{\displaystyle{\sum_{l=0}^{p-1}
\beta_l}} & \mbox{ with } & \beta_n = \frac{C_n^p}{\left(
\varepsilon + IS_n \right) ^2 } \label{WENO3}
\end{eqnarray}
$\varepsilon$ is a small positive number to avoid denominator to
be zero (hereafter we set $\varepsilon = 10^{-6}$) and $IS_n$ is a
measure of the flux function regularity for the $n^{th}$ ENO
stencil candidates. The evaluation of the smoothness measurement
($IS_n$) is based on the undivided-differences and the $IS_n$
formulation can be found in~\cite{jiang_s96}.
It is such that a more regular curve gives a smallest
$IS_n$, and thus a largest weight $\beta_n$.
The $C_n^p$
coefficients are reported in Table~\ref{table_cweno} up to the
order $p=5$, for positive eigenvalues at the cell interface
($\mu^{\pm}_{i+1/2} > 0$). For the negative eigenvalue case, the
WENO coefficients can be obtained by symmetry with respect to the
considered cell interface ($r_{i+1/2}$).

\begin{table}
\centerline{\begin{tabular}{||c|c|c|c|c|c||}
\hline \hline $p$ & $n=0$ & $n=1$ & $n=2$ & $n=3$ & $n=4$ \\
\hline \hline 2 & 1/3 & 2/3 & & & \\
\hline 3 & 1/10 & 6/10 &  3/10 & & \\
\hline 4 & 1/35 & 12/35 &  18/35 & 4/35 & \\
\hline 5 & 1/126 & 10/63 &  10/21 & 20/63 & 5/126 \\
\hline \hline
\end{tabular} }
\caption{The constant coefficients $C_n^p$ of the WENO
reconstruction up to 5$^{th}$-order, for a positive eigenvalue. }
\label{table_cweno}
\end{table}

Using this WENO scheme~(\ref{WENO1}-\ref{WENO3}), the Euler flux
derivative is estimated with a $(2p-1)^{th}$-order of accuracy at
best (in regular regions). Moreover, let us underline that,
if the solution is regular enough, the WENO procedure
recovers
a
high-order {\em centered} scheme, which is of course non-dissipative.

If the length of the stencil is $p=3$ - which is
our case, the order of the scheme in regular regions
is then 5. However, as the order of the scheme
drops to 1 in shock regions - as for any other
scheme - it is not interesting to increase too
much the order in regular regions.
As the scheme is less dissipative with $p=4$
in the non regular regions, this could even
make the results worse (indeed, in our case,
$p=4$ gave less satisfying results), since
spurious oscillations may occur.

\vskip 0.2cm
\noindent
{\bf c) Boundary conditions}
\vskip 0.2cm

To solve the system of equations we need boundary conditions at
the center of the sphere ($r=0$) and at the ceramic surface ($r =
L(t)$). At the center of the sphere, since the velocity of the
fluid is an antisymmetric quantity, the fluid is at rest
($\displaystyle{\left. u \right| _{r=0} = 0 }$). Moreover, as the
pressure and the density are symmetric quantities, their radial
derivatives are set to zero ($\displaystyle{\left.
\frac{\partial}{\partial r} (P, \rho) \right| _{r=0} = 0}$). The
singular behavior of the source terms involves a specific
treatment at the sphere center. While the velocity tends to zero
when the radius tends to zero, the ratio $\displaystyle{\frac{\rho
u}{r}}$ tends to a limit which has to be determined though
$\displaystyle{\frac{\rho u^2}{r}}$ tends to zero. As
$\displaystyle{\frac{\rho u}{r}}$ is a symmetric quantity, its
limit (reached at $r=0$) is calculated by writing that its radial
derivative vanishes at the sphere center by using a second order
upwind difference:
\begin{eqnarray}
\left. \frac{\partial}{\partial r} \left( \frac{\rho u}{r} \right)
\right| _{r=0} & = & \frac{1}{2 r} \left[ -3 \left( \frac{\rho u}{r}
\right) _{i=0} + 4 \left( \frac{\rho u}{r} \right) _{i=1} -
\left( \frac{\rho u}{r} \right) _{i=2} \right]\nonumber \\
& = & 0.
\end{eqnarray}

At the ceramic surface, one has to impose a condition
corresponding to the motion of the ceramics.
In our case, as the displacement of the
sphere does not reach a sonic velocity,
the two eigenvalues
$\mu^{+}$ and $\mu^{-}$ are of opposite sign.
This means that the two informations necessary
to determine the two unknowns $\rho$
and $u$ come from opposite directions.
In particular, at the boundary, one comes from
the interior of the domain and the other from
the outside.
Thus, it is possible to prescribe one of the variable,
and the other one will be determined by
an upwind scheme, i.e. a scheme asymmetric towards
the inner domain.
In the present problem, it is much more natural to prescribe
the velocity at the ceramic surface:
\begin{equation}
u \left( \RL (t) , t \right) = -  \omega \, \Delta x_0 \, \sin
(\omega t) \, \left[ 1 - \exp(-t/1.5) \right] .
\label{ceramic_spedd}
\end{equation}
$\Delta x_0 $ is the amplitude of the displacement of the ceramic
and $\omega$ is the ceramic pulsation. These two parameters are,
of course, prescribed. The exponential term represents the
response time of the transducer (see Sec.~\ref{sect_model}).

\subsubsection{Numerical results}

\begin{figure}
\centerline{\includegraphics[width=8.5cm]{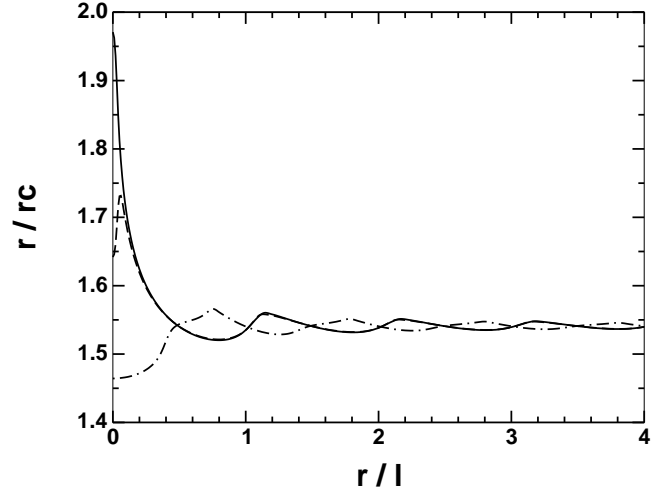}}
\caption{\em Density profile for an oscillation $\DELTAX = 7$ nm,
at different times, corresponding to the maximal (solid line),
minimal (dot dashed line) focal pressure, and to a sharp
front arriving to the center (dashed line) just before
the pressure maximum.
Non linearities are becoming very important.
The simulation was done with $100$
mesh points per wavelength,
starting with a static density $\RHOST = \rho_0$.
}
\label{fignonlinearprof}
\end{figure}

\begin{figure}
\centerline{\includegraphics[width=8.0cm]{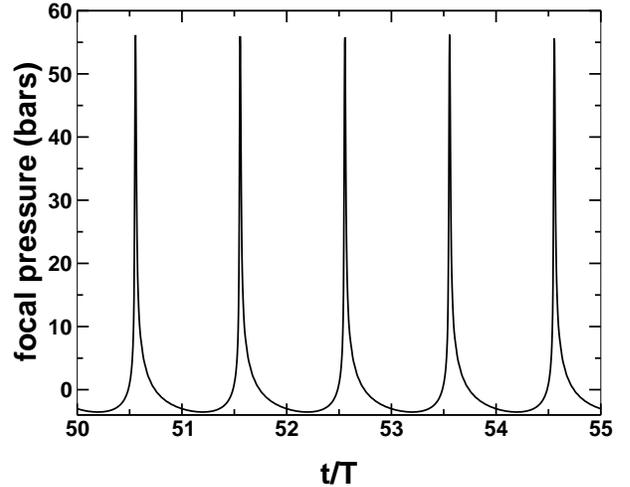}}
\caption{\em Focal pressure for an oscillation $\DELTAX = 7$ nm.
Non linearities are becoming very important.
The simulation was done with $100$
mesh points per wavelength,
starting with a static density $\RHOST = \rho_0$.
}
\label{fignonlinearpres}
\end{figure}

Of course, all the results presented in section
\ref{sect_focal} can be reproduced with our WENO scheme.
Besides, one can increase further the oscillation
amplitude.
For $\DELTAX = 7\,\mathrm{nm}$, 
the reduced shock length (from the center)
is equal to $0.043$, which is now larger than
the spatial step $\delta r/\lambda = 10^{-2}$.
One clearly sees the formation of sharp fronts
(Fig.~\ref{fignonlinearprof}).
At the center, the positive swings of the pressure
are now very sharp compared to the negative ones
(Fig.~\ref{fignonlinearpres}).
The shape of the negative swings is also clearly asymmetric
in time.

\begin{figure}
\centerline{\includegraphics[width=8.0cm]{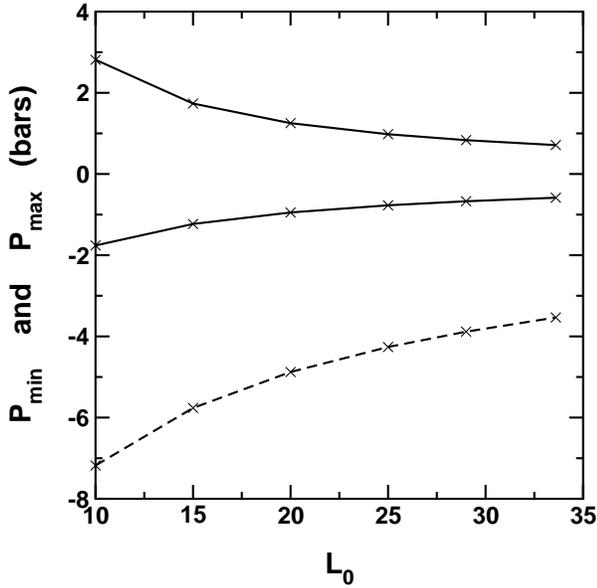}}
\caption{\em Minimal and maximal focal pressure,
as a function of the cell radius.
The product $\RL_\mathrm{red} \times \Delta x_\mathrm{red}$ is constant
for each curve.
Solid lines (resp. dashed lines) correspond to an experimental oscillation
amplitude $\Delta x_\mathrm{exp} = 0.7$ nm (resp. $7$ nm).
One of the curves ($P_{max}$ for $7$ nm) is not
visible, as it varies from 28000 to 106 bars!
In all cases, the simulation was done with about 350
points per wavelength.
}
\label{fig_rdx}
\end{figure}

It would be tempting, in order to reduce the
computational effort, to assume that non-linearities
play a role only in the last wavelengths.
Then, one could simulate a reduced box with
radius $\RL_\mathrm{red}$, and take as an input condition :
\begin{equation}
\Delta x_\mathrm{red} = \Delta x_\mathrm{exp} {\RL_\mathrm{exp} \over
\RL_\mathrm{red}}
\end{equation}
where $\Delta x_\mathrm{exp}$ is the experimental oscillation
amplitude of the transducer, and $\RL_\mathrm{exp}$
its radius.
Actually, this is fine as long as non-linearities
do not play a role at all, i.e. as long as the focal
pressure is sinusoidal.
In all other cases, as this is shown on Fig.~\ref{fig_rdx}, non-linearities are built through
the whole propagation process, and one cannot neglect
them even far from the center, without modifying
the focal pressure.
This could also be seen in the shock length analytic
expression (\ref{rchoc}), which is directly proportional
to the cell radius.
In Fig. \ref{fig_rdx}, we have performed simulations
with various cell radii, corresponding to the same
experimental oscillation amplitude on the ceramic.
As the cell radius
increases, both the positive and negative pressure
swings decrease (in absolute values). Non-linearities
make it more difficult to reach extreme values.
Thus in all the simulations presented in this paper,
we have simulated the whole experimental cell,
with radius $8$ mm.

We have seen on Fig.~\ref{fignonlinearpres}
that positive
pressure peaks can reach tremendously high values.
Of course one may wonder whether this is physical.
The obvious answer is no - but it is worth to
discuss this point in some details.

\begin{figure}
\centerline{\includegraphics[width=8.0cm]{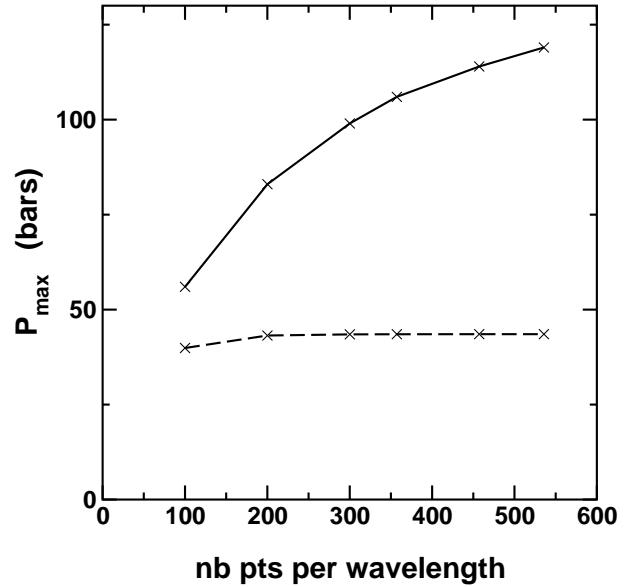}}
\caption{\em Maximum of the focal pressure (solid line)
as a function
of the number of points per wavelength, for an oscillation
amplitude $\Delta x_0 = 7.0$ nm.
For comparison, the dashed line gives the pressure
after a spatial average weighted by a Gaussian with
waist $7 \mu m$.
}
\label{fig_pmaxdvgce}
\end{figure}

\begin{figure}
\centerline{\includegraphics[width=8.0cm]{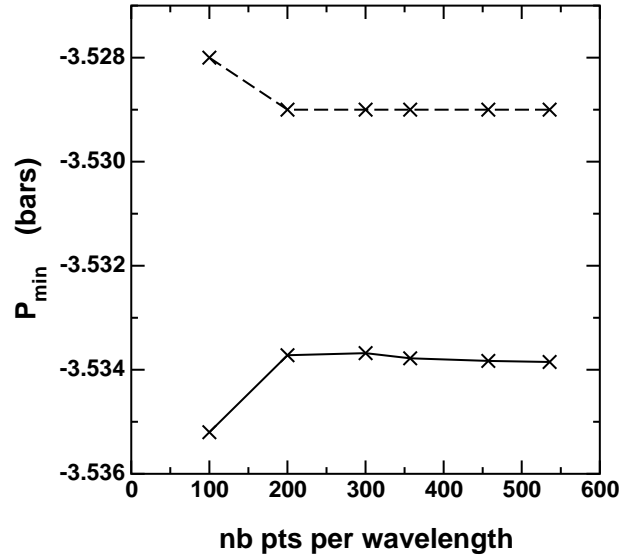}}
\caption{\em Minimum of the focal pressure (solid line)
as a function
of the number of points per wavelength, for an oscillation
amplitude $\Delta x_0 = 7.0$ nm.
For comparison, the dashed line gives the pressure
after a spatial average weighted by a Gaussian with
waist $7 \mu m$.
}
\label{fig_pmindvgce}
\end{figure}

The solution of the problem as we defined it in
Section \ref{sect_model} becomes singular when the shock
reaches the center of the sphere, and positive peaks
actually diverge in the simulations as the spatial
discretisation step $\delta r$ is decreased
(see Fig.~\ref{fig_pmaxdvgce}).
For each fixed $\delta r$, one still finds a finite
focal pressure, as it is defined as an average over
the central cell - with radius $\delta r / 2$.
We check on Fig.~\ref{fig_pmindvgce}
that the minimal pressure does converge
for a decreasing $\delta r$.

\begin{figure}
\centerline{\includegraphics[width=8.0cm]{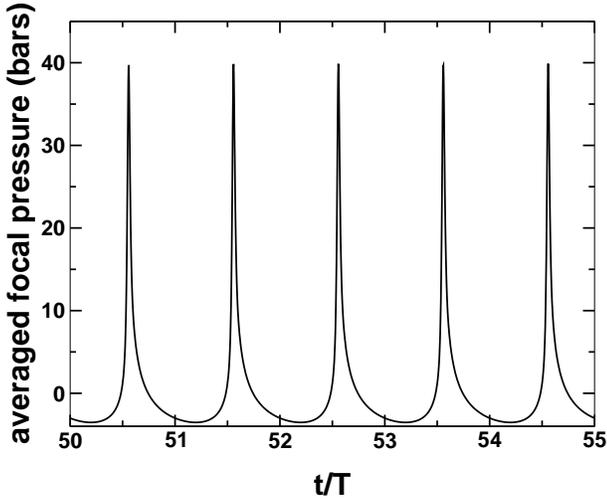}}
\caption{\em Averaged focal pressure for an oscillation $\DELTAX = 7$ nm.
The spatial average is weighted by a Gaussian
with waist $7 \mu m$.
This is to be compared with Fig.~\protect{\ref{fignonlinearpres}}.
}
\label{fig_gaus}
\end{figure}

A first remark is that this singularity involves only
a very small region around $r=0$ (ultimately, it is singular
only in $r=0$).
In the experiment, one cannot measure
the pressure exactly in $r=0$, but rather
over the whole region reached
by the laser beam (see section \ref{sect_exp}).
The intensity of the laser beam through its cross
section is Gaussian, with a waist (half width)
equal to 7 microns.
In order to take this averaging effect into account
in the simulation, we have also computed
a spatial average value of the pressure weighted by
a Gaussian of waist $7 \mu m$ (Fig.~\ref{fig_gaus}).
Positive pressure peaks are lowered, while the
negative swing is not very sensitive to the averaging
process, as gradients are much weaker than during
the positive swing.
But above all, one now obtains extrema which are
converging as $\delta r \to 0$
(Fig.~\ref{fig_pmaxdvgce}).

But it is not enough to rule out the singularity
by an averaging process.
Actually, the singularity is not expected to hold
as such in a more realistic description.
First, the third Euler equation for energy should
also be taken into account in this regime,
as well as regularization mechanisms (dispersion,
dissipation).
Secondly, in the experiment, we expect diffraction,
and the fact that actually the flux is not zero
at the focal point, to break the geometrical symmetry.
Making quantitative estimates of these effects
is not simple, and is postponed to future work.
Finally, even with averaging,
positive pressure peaks in Fig.~\ref{fig_gaus}
are still so high that
they are far above the solidification pressure
(25.3 bar at T=0). As we shall see in the last section, our recent
experiments show that indeed acoustic waves can trigger crystallization, not only
cavitation~\cite{chavanne_b_c01}.
In our simulations, we are not able to treat the possible crystallization.

As a conclusion, the value of the positive pressure swings
is not expected to be reliable for high amplitudes of the transducer surface. 
But we checked, using different types of
numerical regularization of the shocks
(an example is given in the next section),
that even if it affects the maximal pressure,
it has no effect on the
negative swings, and thus it
is still possible to use our simulations to
draw conclusions for negative pressures.

\begin{figure}
\centerline{\includegraphics[width=8.5cm]{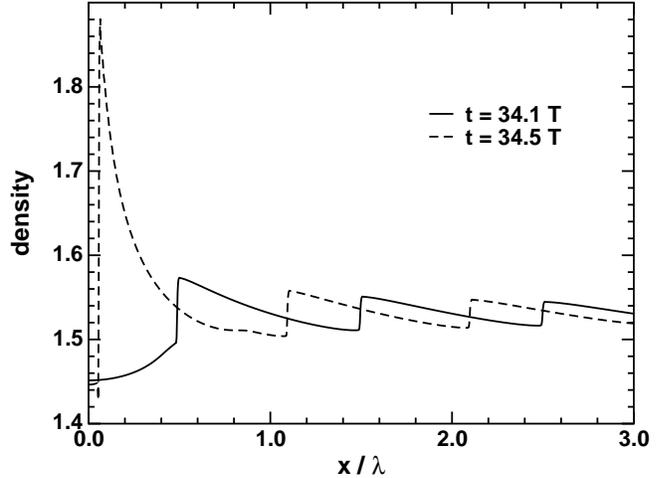}}
\caption{\em Density profile for an oscillation
$\DELTAX = 30$ nm. Shocks form near the focal region.
The simulation was done with about 350
points per wavelength.
}
\label{figchocprof}
\end{figure}

\begin{figure}
\centerline{\includegraphics[width=8.5cm]{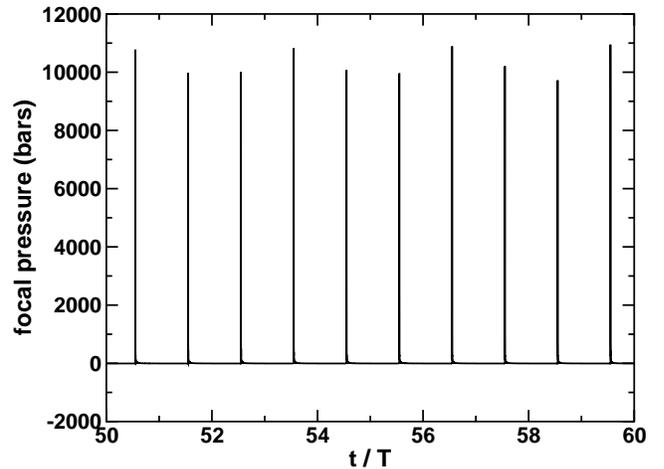}}
\caption{\em Focal pressure for an oscillation $\DELTAX = 30$ nm.
The simulation was done with about 350
points per wavelength.
}
\label{figchocpres}
\end{figure}

In Figs.~\ref{figchocprof} and~\ref{figchocpres},
we show the density profile
and focal pressure obtained for $\DELTAX=30$ nm,
a value which can be reached in cavitation experiments.
Fig.~\ref{figchocprof} illustrates how shocks
are formed when the wave arrives near the focal point.
The amplitude of the shock increases when the shock
itself arrives at the focal point, leading to tremendous
pressure maxima in the simulations (here about 12000 bars).
This is of course unphysical, as we just discussed it,
but the important point is that the minimum
(negative) pressure does not
depend on the value of the maximal pressure.
Thus it is still possible to calculate a minimum pressure value
and compare with experiments.

\begin{figure}
\centerline{\includegraphics[width=8.5cm]{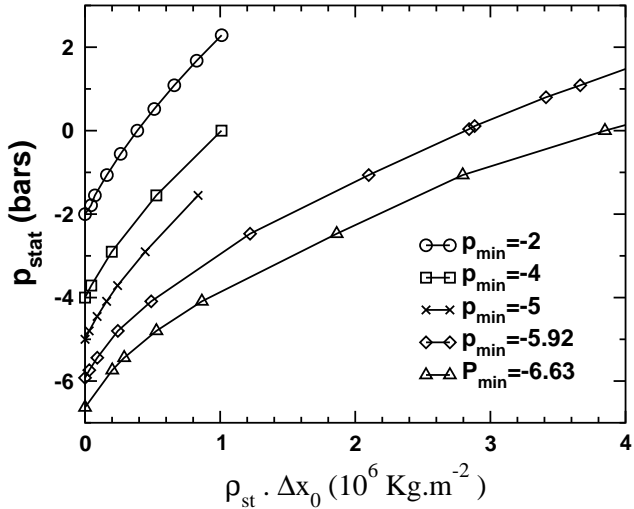}}
\caption{\em Static pressure, as a function of the
oscillation amplitude for which the minimal
focal pressure is equal to a given value $P_\mathrm{min}$,
multiplied by the static density.
Each curve corresponds to a different value of $P_\mathrm{min}$.
}
\label{fig_rhodx0_pstat}
\end{figure}

Fig.~\ref{fig_rhodx0_pstat}
shows the amplitude of the transducer oscillation which is
necessary to reach a given minimal pressure at the center,
starting from a static pressure $P_\mathrm{st}$.
Actually, the variable we plot is rather $P_\mathrm{st}$ as
a function of
$\RHOST . \Delta x_0$, in order to see the departure
from the linear theory
\begin{equation}
P_\mathrm{min} = P_\mathrm{st} + \omega^2 \RL_\mathrm{exp} \RHOST \Delta x_0.
\end{equation}
Our interest in such a curve came from a first version of the
experiment, for which it was not possible to measure the
focal pressure.
Still, we were able to measure the oscillation amplitude necessary
to obtain cavitation for various initial static pressures,
i.e. we could plot a curve such as those of Fig.~\ref{fig_rhodx0_pstat}, for the special case
$P_\mathrm{min} = P_\mathrm{cav}$, the cavitation pressure.
However this could be done only for positive static pressures,
while we were interested in the zero amplitude value of the curve,
for which static pressure and cavitation pressure should be the same.
Thus simulations are useful to determine which kind of extrapolation
should be used for negative static pressure values.

\begin{figure}
\centerline{\includegraphics[width=8.5cm]{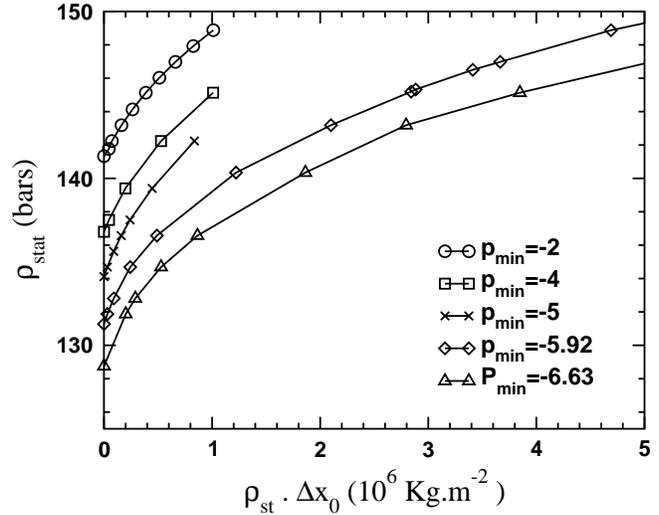}}
\caption{\em Same figure as Fig.~\protect{\ref{fig_rhodx0_pstat}},
except that the static pressure is replaced by the corresponding
static density. The curvature of the function is
enhanced by using this variable.
}
\label{fig_rhodx0_rstat}
\end{figure}

It is
interesting to see that the effect of non-linearities is to bend such
curves in a concave way (Fig.~\ref{fig_rhodx0_pstat}).
In Ref.~\cite{caupin_b01}, the sign of this curvature was
used to show that a linear extrapolation provides an upper bound of the
cavitation pressure.  Whether the shock formation affects the nucleation
mechanism is an open question. If we try to plot the same oscillation
amplitude as a function of the static density in the cell (Fig.~\ref{fig_rhodx0_rstat}),
instead of the
static pressure (Fig.~\ref{fig_rhodx0_pstat}), non-linearities are even more pronounced, as
would have been expected from the equation of state (a concave function
of a concave function is still more concave).

More important for the validation of our theoretical methods is that we
succeeded later in measuring at the focal point the temporal signal
itself. As explained below, this
was done in a quasi-spherical geometry and very good agreement was found
between theory and experiments.

\section{Experiments}

\label{sect_exp}

A hemispherical piezoelectric transducer is held against a clean
glass plate. In
a first approximation, the glass reflects the sound wave so that this is
equivalent to a full spherical geometry. The main interest of the glass plate
is that it allowed us to
measure the instantaneous density at the center from the reflection of light
at the glass/helium interface. Indeed, the reflectance depends on the
refraction index of liquid helium which depends on its density as is well
known from the Clausius-Mossoti relation.

\subsection{Experimental method}

\begin{figure}[ttt]
\centerline{\includegraphics[width=8.0cm]{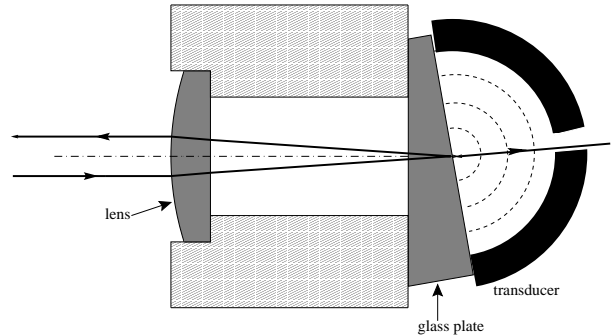}}
\caption{The experimental set-up which is immersed in liquid helium, inside
the experimental cell. At the center of the hemispherical transducer,
the amplitude
  of the sound oscillation is measured from the intensity of the reflected
  light. The transmitted light is used to detect the possible
nucleation of bubbles
  or crystallites.}
\label{fig:optiq}
\end{figure}

\begin{figure}[ttt]
\centerline{\includegraphics[width=8.0cm]{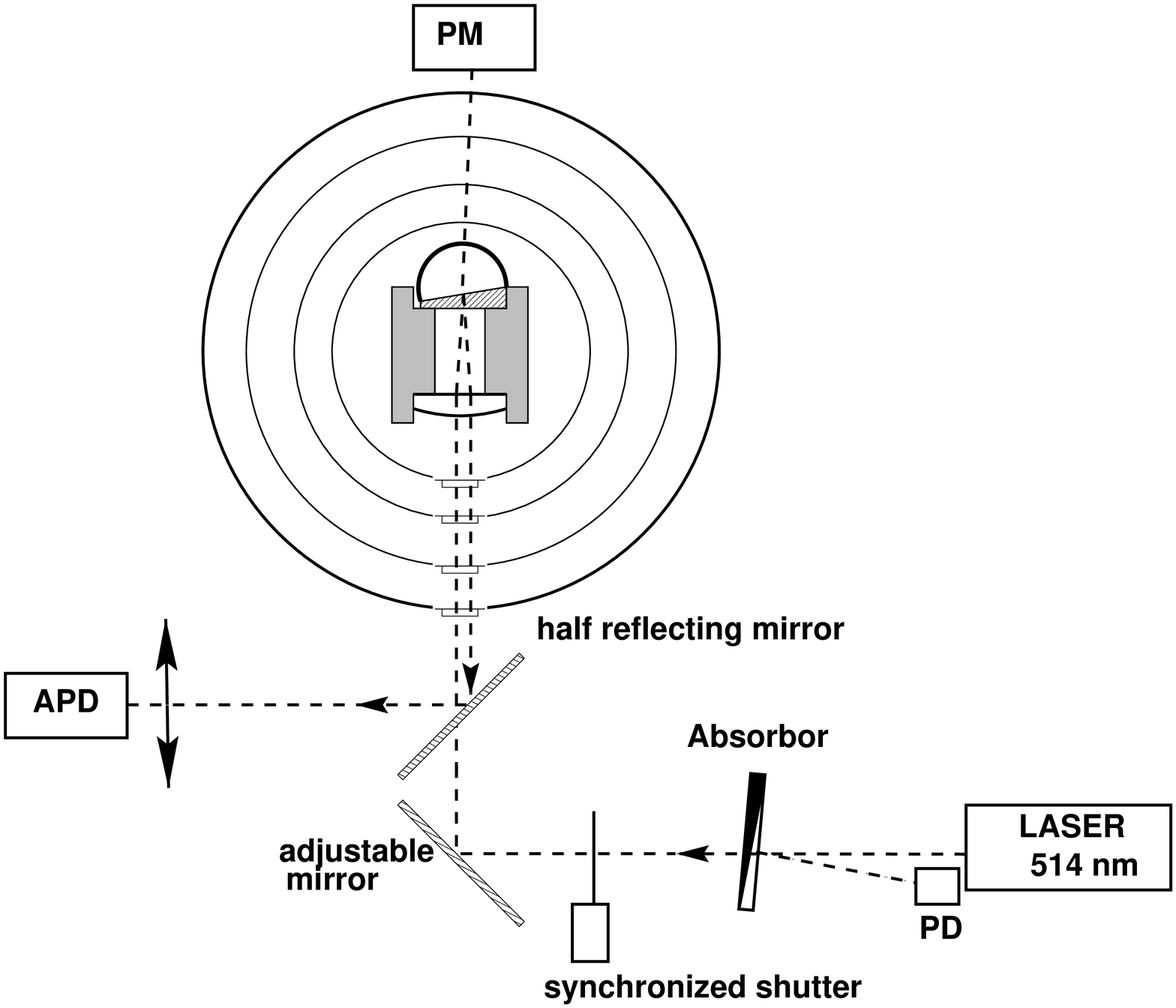}}
\caption{dispositif experimental}
\label{fig:dispo}
\end{figure}

Our experimental method is described in full details
elsewhere~\cite{chavanne_b_c01c,chavanne_b_c02}. Let us only summarize it here. 
The transducer radius is $8\,\mathrm{mm}$ and its thickness is $2\,\mathrm{mm}$. It resonates in a thickness mode at $f$ = 1.019 MHz. At this frequency, it has a minimum impedance $Z$ = 22 Ohm. Its quality factor is $Q$ = 50 $\rm \pm$ 5 when immersed in liquid helium at 25 bar. We usually pulse it with bursts of 6 oscillations. The $300\,\mathrm{cm^3}$ experimental cell is full of liquid helium and attached to the mixing chamber of a dilution
refrigerator. We can run the experiment between 30 mK and 1.5~K, at
static pressures from 0 to 25 bar. Fig.~\ref{fig:optiq} shows our optical setup: a brass piece holds a lens whose focal length is 21.8 mm in liquid helium and a wedged glass plate (20 mm in diameter, $\approx$ 2 mm thick). The transducer is pressed against the plate. All the space inside is filled with liquid He. Thanks to the lens, the radius of the laser waist is reduced from 320 $\rm \mu m$ to 7 $\rm \mu m$. This means that the spatial resolution is about 14 $\rm \mu m$, the diameter of the optical focal region. This is small compared to the size of the acoustic focal region which is set by the acoustic wavelength at 1~MHz: from 240 $\rm \mu m$ at 0 bar to 360 $\rm \mu m$ at 25 bar. The distance of the lens to the glass plate has been carefully adjusted to have the laser focused at the glass/helium interface, on the transducer side. This was checked from the parallelism of the reflected beam. The 2 degrees wedge of the glass plate avoids interferences with reflections on its front face. A 1.7 mm hole in the transducer allows the transmitted light to be analyzed on the other side of the cryostat (see Fig.~\ref{fig:dispo}).

We use a single mode $\rm
Ar^+$ laser. It is operated around 10 mW where its stability is best. In front of the laser is an absorber which reduces the light entering the cell (Fig.~\ref{fig:dispo}). We use the reflection on the front face of the absorber to monitor possible drifts of the laser power, but its stability is better than 0.5 \% per day. For low temperature measurements,
we reduced the dissipation in
the cell with a small
electromechanical shutter; it
was synchronized by the pulse generator so that the cell was illuminated
during 10 msec only, around the arrival time of each acoustic pulse on the
glass plate. Two moving mirrors allow us to translate the laser beam vertically and horizontally. The last mirror is mounted on a rotating plate, so that the angle of incidence of the beam can be adjusted as
well. These rotations correspond to translations of the optical focus in the focal plane of the lens. By successive operations, the laser spot was brought to the center of the acoustic focal region. The final adjustment was obtained by maximizing the modulation of the reflected signal by the acoustic wave.

The transmitted light is collected by a photomultiplier and used to
detect nucleation
events one by one. The light which is reflected at the glass/helium
interface is separated from the incident beam by means of a semi-transparent plate and directed
towards a photodiode. We used either a Hamamatsu C5331-03 avalanche photodiode (the``APD''), for the detection of the ac-modulation of the reflected beam, or a Hamamatsu S1406 silicon photodiode (the``SPD'') for the detection of the dc-component. The output from the photodiodes is digitized with a LeCroy 9344 CM oscilloscope at 1 GS/s with 8-bit resolution (6.5 effective bits due to the clock jitter at 1 GS/s).

\begin{figure}[ttt]
\vspace{-1cm}
\centerline{\includegraphics[width=9cm]{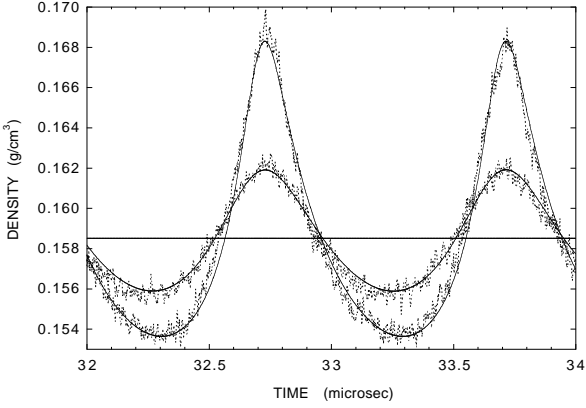}}
\caption{Two recordings of sound wave amplitudes respectively
corresponding to excitation voltages 9.05 and 20.4 V on the transducer. The
static density is $\rho \rm = 0.15851 \: g/cm^3$
corresponding to a static pressure
$P_{stat} = 9.80$ bars (horizontal line).
The asymmetry of the oscillations is well reproduced by the numerical
calculations performed for $\DELTAX = 3$ and $6$ nm.
The numerical focal density (solid lines) is obtained with
a spatial average weighted by a Gaussian with
waist $7 \mu m$.
Simulations are performed with 350 mesh points per wavelength.
}
\label{fig:comp}
\end{figure}

The ac-component of the signal is related to the modulation of the helium
density by the acoustic wave, and it is at most a few percent of the dc part which is related to the static density.
The acoustic transmission into the glass being very small, we have
neglected its effect on the reflected light. In order to achieve a 1\% accuracy on the acoustic wave amplitude, we reduced the noise by averaging on 10000 sound bursts with a repetition rate of 1 to 10 Hz. There are two main sources of noise. The first one is photon noise in the reflected light. For a 5 $\mu$W power, this quantum noise is typically 10 nW, much more than the resolution we need. The other noise source is the oscilloscope jitter
and it has a comparable amplitude. After averaging, the signal is well
enough extracted from the noise as shown on Fig.~\ref{fig:comp}.

\subsection{Calibration}
The intensity of the reflected light is proportional to the normal reflectance R at the glass/helium interface:
\begin{equation}
R=\left (\frac{n_\mathrm{g}-n}{n_\mathrm{g}+n}\right )^2
\end{equation}
where $n_\mathrm{g} = 1.5205$ is the refractive index of glass for 514.5 nm green
light. As for the index n of helium it is given by the Clausius
Mossoti relation:
\begin{equation}
\frac{n^2 - 1}{n^2+2} = \frac{4 \pi \rho \alpha_M}{3M}
\end{equation}
where M = 4.0026 g, $\rho$ is the helium density, and $\alpha_M$ = 0.1245
$\mathrm{cm^3mol^{-1}}$ is the
molar polarizability for the same green light. Note that $\alpha_M$
increases slightly as a function of frequency from its zero frequency
value $\alpha_{M0} = 0.1233 \,\mathrm{cm^3mol^{-1}}$, as explained by
successive authors~\cite{polar}.

We proceed as follows. We first measure the static pressure in
the cell. Knowing the equation of state $P(\rho)$, we obtain the static
density~\cite{maris91}. From the static density, we calculate
the normal reflectance in the absence of modulation by the
wave. This is our
reference. We then measure the ratio of the ac- to the
dc-component of the
reflected light, and obtain the amplitude of the density modulation in the
acoustic wave. Unfortunately, we cannot do this with a single
diode, so that we have to calibrate the ratio of the respective gains of our two
photodiodes. This is achieved in the overlap of their
bandwidths, with the help of an acousto-optic modulator operated
at 200 kHz. We also have to check the linearity of the APD. Its gain is found constant
up to incoming powers of 6 $\mu W$ where it starts decreasing. Most of the time, we use the APD in its linear regime;
otherwise, a small correction is applied.

We finally have the following sources of uncertainties: the static density
can be known within $2$ to $4\,10^{-2} \,\mathrm{kg\,m^{-3}}$. About the same
uncertainty comes from the determination of the base line of the APD signal.
Uncertainties in the gain ratio and in the APD measurement lead to a
1\% uncertainty in the wave
amplitude. We could finally check our calibration by studying
heterogeneous nucleation of bubbles on the glass plate. We observe various
nucleation mechanisms. One of them occurs at saturated vapor pressure
($P_\mathrm{sv} = 0 \,\mathrm{bar}$ in the low temperature limit) where the liquid density is
$145.13 \,\mathrm{kg\,m^{-3}}$. In a series of measurement at a static pressure
$P_\mathrm{st} = 4.30 \,\mathrm{bar}$, we found cavitation at $145.15 \,\mathrm{kg\,m^{-3}}$; in
another series of measurements at $P_\mathrm{st} = 2.95\,\mathrm{bar}$, we found cavitation
at $145.12 \,\mathrm{kg\,m^{-3}}$. This illustrates the final uncertainty in our
measurements.

\subsection{Comparison with calculations}

Fig.~\ref{fig:comp} shows two recordings obtained at 0.1K with
respective excitation
amplitudes of 9.05 and 20.4 V on the transducer. In the cell, the static
pressure is 9.80 bar, corresponding to a static density $\rho = 158.51 \,\mathrm{kg\,m^{-3}}$. The density oscillation is found asymmetric at large amplitude:
negative swings are broader with a smaller amplitude than positive swings. Moreover, the negative swings are not symmetric in time. We have chosen this recording at intermediate pressure and moderate amplitude because the signal shape is not modified by any nucleation
of crystals or bubbles.

We can compare the experimental recordings with the numerical calculations
described in sections \ref{sect_simul} and \ref{sect_weno}.
Simulations are performed with the same static pressure as
in the experiment.
Then there is only one free parameter in the simulation
(the oscillation amplitude $\DELTAX$)
in order to adjust both the amplitude
and the shape of the signal. The adjustment with the experimental signal
is made only on the central oscillation of the latter.
Indeed, in the experiment, the transducer is excited with
an electrical burst of six oscillations. Since the transducer has a finite
quality factor $Q \approx $ 50, the amplitude of the sound wave increases
during six periods and slowly decreases afterwards. The
numerical result is used in the steady regime, after the initial
transient. As
can be seen, we find a very good agreement for numerical oscillation
amplitudes $\DELTAX = 3$ and $6$ nm: both the asymmetry with respect
to the horizontal axis and the asymmetry in time are well reproduced by
the calculation.

\begin{figure}[ttt]
\centerline{\includegraphics[width=8.5cm]{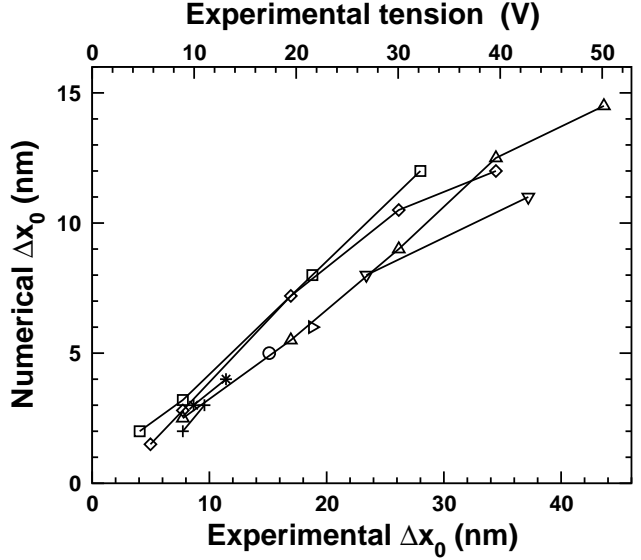}}
\caption{Oscillation amplitude $\DELTAX$ for which the simulation
fits the experimental signal, as a function of the experimental
applied voltage. These results are obtained for different static
pressures, i.e. 110 mbars ($\star$), 800 mbars ($\circ$),
4.3 ($\square$), 5.1 ($\diamondsuit$), 9.8 ($\triangle$),
15.4 ($\triangledown$), 22 ($\triangleright$), and
25.3 ($+$) bars.
We also indicate as a second x-axis the estimate for the
oscillation amplitude given by (\protect{\ref{eq_estimate}})
which is obtained from independent physical arguments.}
\label{fig_dx_dx}
\end{figure}

We make similar adjustments with several recordings, for different
oscillation amplitudes and different static pressures. Our results
are summarized on Fig.~\ref{fig_dx_dx}.
For each adjustment, the numerical amplitude obtained by fitting
the central oscillation is associated to the experimental
voltage applied to the transducer.
On Fig.~\ref{fig_dx_dx}, we also indicate as a second x-axis
the estimate given by (\ref{eq_estimate})
for the oscillation amplitude $\ZETA$.
As this estimate is obtained independently from the simulations,
as this will be detailed below, we refer to it as the experimental
oscillation amplitude in the figure.
If the estimate and the numerical adjustment were in perfect agreement,
one would expect a slope equal to one in Fig.~\ref{fig_dx_dx}.
Actually, we find that both methods give different results,
and this will be discussed below.
The adjustment between experimental signals and numerical simulations
yields the following calibration~:
\begin{equation}
\frac{\ZETA}{V} = 0.30 \pm 0.02 \,\mathrm{nm\,V^{-1}}.
\label{eq:calib}
\end{equation}

It is interesting to compare this value with an estimate from the measurement
of the electrical characteristics of the ceramic. Let us summarize the derivation of this estimate, which is given in Ref.~\cite{caupin_b01}.

 Indeed, the quality
factor is simply related to the ratio of the acoustic energy $E_\mathrm{ac}$
which is stored during one period to the average
dissipated power $V^2/(2Z)$:
\begin{equation}
Q = 4\pi f \frac{E_\mathrm{ac}\,Z}{V^2} \: ,
\end{equation}
and the acoustic energy can be evaluated as follows. Let us call $R$ the mean
radius of the transducer and $2e$ its thickness. For a resonance in a
thickness mode, one can assume that the sound wave inside the transducer
is a spherical wave with an amplitude proportional to $1/r\: \sin
[(k(r-R)] \sin (\omega t)$
where the wavevector $k = \pi /2e$ and $\omega = 2 \pi f$. For a spherical
wave, the elastic displacement inside the transducer is

\begin{equation}
u = u_0 \frac{R}{r} \sin [k(r-R)] \sin (\omega t).
\end{equation}

Since $E_\mathrm{ac}$
is twice the average kinetic energy in the transducer~\cite{caupin_b01}
and the local velocity is simply the time derivative of $u$, one
can integrate over the thickness and write:

\begin{eqnarray}
E_\mathrm{ac} = \pi \rho_\mathrm{t} \omega^2u_0^2 R^2 e
\end{eqnarray}
where $\rho_\mathrm{t}$ is the transducer density. After expressing $E_\mathrm{ac}$
in terms of the
maximum displacement $\ZETA = u_0 R/(R-e)$ of the inner surface, we obtain

\begin{equation}
E_\mathrm{ac} = \frac{M}{4} \omega^2 \ZETA^2 \frac{(R-e)^2}{R^2+\frac{e^2}{3}}
\end{equation}
where $M = 7.5 10^{-3}\,\mathrm{kg}$ is the mass of our transducer.
Note that the last factor was forgotten in Ref.~\cite{caupin_b01}.
 Finally, we can express the
displacement $\ZETA$ as a function of the applied voltage:

\begin{equation}
\ZETA = V \left ( \frac{2Q}{\omega^3 M Z} \right )^{1/2} \frac{R-e}{R} \sqrt{1+\frac{e^2}{3 R^2}}
\label{eq:res}
\end{equation}

The above equation leads to $\ZETA / V \rm = 1.7\,\mathrm{nm\,V^{-1}}$. However, this value would correspond to an excitation
with long bursts, when the stored energy saturates. Since we excite it with
bursts of six oscillations only, and since the phase is such that they
 start with positive swings, the maximum pressure is reached (6 +
1/4) periods after time zero, and the
displacement to be considered in our case is
\begin{equation}
\ZETA = \ZETA_{\infty} \left [1-\exp(-12\pi/Q)\right ]\left
[\exp(-\pi/2Q)\right]
\end{equation}
We find $\ZETA = 0.513 \,\ZETA_{\infty}$ with $\ZETA_{\infty}$ given by
Eq.~\ref{eq:res}, and our final prediction is
\begin{equation}
\frac{\ZETA}{V} = 0.87\,\mathrm{nm\,V^{-1}}
\label{eq_estimate}
\end{equation}

The above value has the right order of magnitude but it
is three times more than given by the fit of our numerical
calculations (Eq.\ref{eq:calib}). There are several assumptions in
the above analysis which can
be claimed as responsible for this discrepancy. We list
them starting with those that we expect to be the most
relevant:\\
a - the resonance in a thickness mode may be coupled to flexion modes,
 in which case the efficiency of the transducer can easily be
reduced.\\
b - there is a small hole in the center of the transducer which allows the
transmitted light to be analyzed.\\
c - the sound wave in the transducer cannot be strictly spherical,
since there must be edge effects near its free equator.\\
d - the reflexion by the glass plate is not perfect so that our closed
hemispherical geometry is not strictly equivalent to a full spherical
geometry. Once more this should reduce the efficiency of the transducer.\\
e - some of the emitted energy is lost in the various pieces which hold it
in the cell.\\
f - there is also some uncertainty of order 10 $\%$ in the measurement of
$Z$ and $Q$.

As a result, we consider the value $0.3\,\mathrm{nm\,V^{-1}}$ as a very useful
calibration of the efficiency of our transducer, in qualitative agreement
with a simple estimate.

\section{Conclusions and perspectives}

In this article, we have presented analytic and numerical calculations of the
focusing of a spherical acoustic wave. We have shown that shocks are
generated in
this geometry and we have obtained an analytic estimate for the shock length
based on the characteristics method. Then, in order to perform full
numerical simulations of the focussing process, we have used a WENO scheme
to treat shocks. We
then showed that our method is validated by a comparison with
experimental measurements in a quasi-spherical geometry. We have measured a wave distortion which is well reproduced by our
calculation and the analysis of its dependence on the excitation amplitude
has led us to a very useful calibration of the efficiency of our
transducers.

We consider this work as a first step only, and we plan to
extend it to a hemispherical geometry for two important reasons. Indeed, in
order to study the homogeneous nucleation of bubbles in stretched fluids
or that of crystals in pressurized fluids, we need to eliminate the effect of
walls. This is achieved by using hemispherical transducers which focus
acoustic waves away from any walls~\cite{lambare98,caupin_b01}. In such experiments,
since we have no probe in the acoustic focal region where nucleation
takes place, there is a difficult problem of calibration of the sound
amplitude, for which any reliable calculations would be very useful. Of
course, the calculation in a hemispherical geometry is much more difficult
because it is two-dimensional (it depends on both the radial distance and
the polar angle). Now that the method is known for the treatment of
shocks, the 2-D calculations should be tried. Furthermore, we have
estimated the amplitude of non-linear effects in the hemispherical
geometry~\cite{caupin_b01}, and found them much smaller than in the spherical
geometry, though lower pressures seem to be reached.
This is interesting in itself and should be tested numerically.
One physical explanation could be that the local condition at the center
is different: by symmetry, the spherical geometry imposes that
 the center is a node for the fluid
velocity.
In the hemispherical geometry,
there is no reason why it should be so.
On the contrary, the sound wave could even create
a flow at the center with non vanishing averaged value.
This phenomenon is known in the literature as
acoustic streaming.
This symmetry difference might lead to a different amplitude for the
non-linear effects. It would be very interesting to study this
phenomenon numerically.

Another direction of research would deal with the interaction
between shocks and nucleation. Until now, all theories predicting
the nucleation threshold completely
ignore the presence of very steep gradients. This is not
necessarily justified.

{\bf Acknowledgments:}
We are grateful to H. Lambar\'e for his contribution to the calculations in
their early stage. C.A. would like to thank Frederic Coquel for
interesting discussions.

\vskip 1 true cm
{\sc \bf Appendix : theoretical prediction of the
shock length by the characteristics method}
\vskip 1 true cm

\noindent
{\bf Appendix~A: Characteristic equations}

The system of Euler equations (\ref{eqeuler}) is of the form
\begin{equation}
\dert v_i + A_{ij} \derr v_j = b_i,
\label{eq_syst}
\end{equation}
with ${\bf v} \equiv (\rho, u)$
and ${\bf b} = (-\theta \rho u/r,0)$.
The matrix
$$ A = \left( \begin{array}{cc}
u & \rho \\
{c_\mathrm{s}^2\over \rho} & u
\end{array}
\right)
$$
has two eigenvalues
$\mu_+ = u + c_\mathrm{s}$ and $\mu_- = u - c_\mathrm{s}$
associated with the left eigenvectors
\begin{eqnarray}
{\bf l}_+ & = & \pmatrix{ c_\mathrm{s} / \rho \cr 1\cr },\nonumber \\
{\bf l}_- & = & \pmatrix{ - c_\mathrm{s} / \rho \cr 1\cr }. \label{foo}
\end{eqnarray}
If we apply ${\bf l}_k$ on the left of equation (\ref{eq_syst}),
we obtain
\begin{equation}
{\bf l}_k . \left[ \frac{d{\bf v}}{dt} - {\bf b} \right] = 0,
\label{eq_carac}
\end{equation}
where
\begin{equation}
\frac{d}{dt} = \dert + \mu_k \derr. \nonumber
\end{equation}
Thus the derivative $d/dt$ is taken
along a curve $r=x(t)$ with slope $dx/dt = \mu_k$ everywhere
($\mu_k$ itself being a function of $r$ and $t$ via $v$).
By definition, this curve is called a k-th characteristic
and is denoted by $\CCK$. There is a whole family of
$\CCK$ characteristics, covering the whole space,
each curve being determined, for example,
by the initial conditions.

The leftmost term of equation (\ref{eq_carac}) can be integrated
and the equation becomes
\begin{equation}
\frac{d}{dt}\left[ C_\mathrm{o}(\rho - \ln \rho) \pm u\right]
+ \theta \frac{C_\mathrm{o}(\rho -1) u}{r} = 0,
\label{eq_charapp}
\end{equation}
where again the time-derivative is taken along a
characteristic curve.

We call Rieman invariants the quantities
\begin{equation}
\INVPM \equiv C_\mathrm{o}(\rho - \ln \rho) \pm u
\end{equation}
appearing in eq. (\ref{eq_charapp}).

\vskip 1cm
\noindent
{\bf Appendix~B: Lower bound for the shock length}
\vskip 1cm

Considering the geometry and notations of Sec.~\ref{sect_analytic},
we calculate at what time
the first characteristic $\CCD_0$ emitted by the piston
at $t=0$
is cut by another characteristic $\CCD_i$.
This is the signature of a shock forming.
The calculation assumes that it is
the first shock ever formed in the cell.

As $\CCD_0$ is first cut by characteristics $\CCD$
emitted at early times, and almost parallel to $\CCD_0$,
we perform the following change of variable.
Instead of $r$, the location
of any point will be given by its distance $\eta$ to a point
moving on the characteristic $\CCD_0$, taken at the same time :
\begin{equation}
\eta = \RL_0 - c_\mathrm{st}t - r.
\end{equation}

The characteristic equations (\ref{eq_charapp})
have to be written in the new coordinates $(\eta, t)$.
Besides, we would like to eliminate $u$ and $\rho$ from the equations
so that the only remaining unknowns would be the Rieman
invariants $\INVUN$ and $\INVD$.
Following~\cite{greenspan_n93}, we have $d\eta /dt = - (u\pm c_\mathrm{s} + c_\mathrm{st})$
and $u = (\INVUN - \INVD)/2$. But unlike Ref.~\cite{greenspan_n93},
the density $\rho$ -and thus the sound velocity-
cannot be expressed in a simple way in terms of $\INVUN$ and $\INVD$.
Indeed, we would need to invert the relation
\begin{equation}
\INVUN + \INVD = 2 C_\mathrm{o} (\rho - \ln \rho) \equiv 2 I(\rho).
\end{equation}
At this stage, we will only assume that the inversion
can be performed and write
\begin{equation}
\rho = \IOP^{-1}\left( \frac{\INVUN + \INVD}{2}\right).
\label{eq_inv}
\end{equation}
The characteristic equations now read
\begin{eqnarray}
\dert \INVUN -\! \left\{ \frac{\INVUN-\INVD}{2} + C_\mathrm{o} \left[ \IOP^{-1}
\left( \frac{\INVUN + \INVD}{2}\right)-1\right] + c_\mathrm{st}\right\} \dereta \INVUN
\nonumber \\
+ \frac{2}{\RL - c_\mathrm{st}t - \eta} \left(\frac{\INVUN-\INVD}{2}\right) C_\mathrm{o}
\left[ \IOP^{-1}\left( \frac{\INVUN + \INVD}{2}\right)-1\right] = 0
\nonumber \\
\label{eqI1}
 \\
\dert \INVD - \! \left\{ \frac{\INVUN-\INVD}{2} - C_\mathrm{o} \left[ \IOP^{-1}
\left( \frac{\INVUN + \INVD}{2}\right)-1\right] + c_\mathrm{st}\right\} \dereta \INVD
\nonumber \\
+ \frac{2}{\RL - c_\mathrm{st}t - \eta} \left(\frac{\INVUN-\INVD}{2}\right) C_\mathrm{o}
\left[ \IOP^{-1}\left( \frac{\INVUN + \INVD}{2}\right)-1\right] = 0
\nonumber \\
\label{eqI2}
\end{eqnarray}
We search for a solution under the form
\begin{eqnarray}
\INVUN (\eta,t) & = & \sum_{m=0}^\infty \INVUN^{(m)}(t) \eta^m,
\nonumber\\
\INVD (\eta,t) & = & \sum_{m=0}^\infty \INVD^{(m)}(t) \eta^m.
\label{eq_expeta}
\end{eqnarray}
We choose the lowest order terms $\INVUN^{(0)}$ and $\INVD^{(0)}$
equal to their value in the fluid at rest
 $\INVUN^{(0)} = \INVD^{(0)} = C_\mathrm{o}(\RHOST -\ln \RHOST)$.
Then if we write equations (\ref{eqI1}-\ref{eqI2}) at lowest order,
all the terms in the second equation vanish.
In the first equation, both the first and last terms disappear.
The remaining term leads to
\begin{equation}
\INVUN^{(1)} = 0.
\end{equation}
This is not very surprising, as $\INVUN$ corresponds to the
characteristics moving from the fluid at rest into the perturbed
region.
Taking (\ref{eqI2}) to the next order,
we obtain an equation for $\INVD^{(1)}$
\begin{equation}
\frac{d \INVD^{(1)}}{dt} + \frac{1}{t - \RL_0 / c_\mathrm{st}} \INVD^{(1)}
+ \KKK \left[ \INVD^{(1)}\right]^2
= 0,
\end{equation}
where
\begin{equation}
\KKK \equiv \frac{1}{2}\;\left( \frac{2\RHOST-1}{\RHOST-1}
\right).
\end{equation}
This equation can be solved using a change of
variables $\overline{\INVD} \equiv 1/\INVD^{(1)}$.
The solution yields
\begin{equation}
\INVD^{(1)}(\tau) = \frac{\INVD^{(1)}(0)}
{(1-\tau) \left[ 1-\KKK \frac{\RL_0}{c_\mathrm{st}}\;\INVD^{(1)}(0)\; \ln(1-\tau)
\right]},
\label{eq_psi1}
\end{equation}
with
\begin{equation}
\tau \equiv \frac{c_\mathrm{st}}{\RL_0} t.
\end{equation}
We have now to determine the initial value $\INVD^{(1)}(0)$.
It refers to small $t$, rather than $t$ exactly equal to zero.

To determine its value, we calculate for small time $t$
the variation of $\INVD$
between point $A = (r,t) = (\RL_0 - c_\mathrm{st}t,t)$ which sits
on the first characteristic $\CCD_0$,
and point $B = (r,t) = (r_p(t),t)$ where $r_p(t)$ is
the location of the piston at time $t$.

At $A$, the fluid is at rest. We have $\INVD = \INVD^{(0)}$
and $\eta = 0$.

At $B$, for $t$ small, $r \simeq \RL_0$ and thus $\eta \simeq - c_\mathrm{st}t$.
Besides, $\INVD = C_\mathrm{o}(\rho -\ln \rho) - u$ where
$u$ is equal to the velocity of the piston
$v_p(t) = -\Delta v_0 \sin (\omega t)
\simeq -\Delta v_0 \omega t$.
If we expand $\rho(r_p(t),t) = \RHOST + \alpha t$ (where $\alpha$
is unknown), then
replacing into $\INVD$ and expanding in $t$ yields
\begin{eqnarray}
\INVD & = & C_\mathrm{o}(\RHOST -\ln \RHOST) + C_\mathrm{o}\alpha t
- C_\mathrm{o} \frac{\alpha}{\RHOST} t + \Delta v_0 \omega t \nonumber\\
& = & \INVD^{(0)} + \left( {c_\mathrm{st} \over \RHOST } \alpha
+ \Delta v_0 \omega \right) t. \nonumber
\end{eqnarray}
Comparing this relation with the expansion
\begin{equation}
\INVD = \INVD^{(0)} + \eta \INVD^{(1)}
\simeq \INVD^{(0)} - c_\mathrm{st} \INVD^{(1)} t,
\end{equation}
gives
\begin{equation}
\INVD^{(1)} = - {\alpha \over \RHOST } - {\Delta v_0 \over c_\mathrm{st}}
\omega.
\end{equation}
On the other hand, an expansion of (\ref{eq_inv}) in powers of
$\eta$ gives
\begin{equation}
\alpha t = \RHOST \RHOST = {c_\mathrm{st} \over \RHOST }
\left( \eta {\INVD^{(1)} \over 2} \right) t
\end{equation}
The elimination of $\alpha$ between the above two equations
yields
\begin{equation}
\INVD^{(1)}(t=0) = - 2 {\Delta v_0 \omega \over c_\mathrm{st}}.
\end{equation}

The time at which $\INVD^{(1)}$ becomes infinite
(see equation \ref{eq_psi1}) gives
an upper bound $t_\mathrm{shock}$ for shock formation.
It is only an upper bound because some other terms of the
expansion (\ref{eq_expeta}) in $\eta$
may explode before $\INVD^{(1)}$. We find
\begin{equation}
t_\mathrm{shock} \le \frac{\RL_0}{c_\mathrm{st}} \left\{ 1 -
\exp\left[ -\frac{c_\mathrm{st}^2}{2 \RL_0 \omega \Delta v_0}
\frac{\RHOST-1}{\RHOST-\frac{1}{2}} \right] \right\}
\le \frac{\RL_0}{c_\mathrm{st}}.
\end{equation}
As the corresponding shock distance $r_\mathrm{shock}$
is measured from the center of the
sphere, a lower bound for $r_\mathrm{shock}$ is
\begin{equation}
\RL_0 \ge r_\mathrm{shock} \ge \RL_0 \exp\left[
 -\frac{c_\mathrm{st}^2}{2 \RL_0 \omega \Delta v_0}
\frac{\RHOST-1}{\RHOST-\frac{1}{2}} \right] > 0.
\label{rchocapp}
\end{equation}

All the above calculations are valid for small $\eta$,
i.e. only for characteristics not too far from $\CCD_0$.
These are the characteristics emitted by the initial motion
of the piston.

\vskip 1cm
\noindent


\begin{thebibliography}{100}

\bibitem{lambare98}
H.~Lambar\'e, P.~Roche, S.~Balibar, H.J. Maris, O.A. Andreeva, C.~Guthman, K.O.
 Keshishev, and E.~Rolley, Eur. Phys. J. B \textbf{2}, 381 (1998)

\bibitem{caupin_b01}
F.~Caupin and S.~Balibar, Phys. Rev. B \textbf{64}, 064507 (2001)

\bibitem{chavanne_b_c01}
X.~Chavanne, S.~Balibar, and F.~Caupin, Phys. Rev. Lett. \textbf{86}, 5506 (2001)

\bibitem{sirotyuk}
M.G. Sirotyuk, Sov. Phys. Acoustics \textbf{8}, 165 (1962)

\bibitem{roy_m_a90}
R.A. Roy, S.I. Madanshetty, and R.E. Apfel, J. Acoust. Soc. Am. \textbf{87}, 2451 (1990).

\bibitem{nemirovskii90}
S.K. Nemirovskii, Sov. Phys. Usp. \textbf{33}, 429 (1990)

\bibitem{greenspan_n93}
H.P. Greenspan and A.~Nadim, Phys. Fluids A \textbf{5}, 1065 (1993)

\bibitem{maris91}
H.J. Maris, Phys. Rev. Lett. \textbf{66}, 45 (1991)

\bibitem{chavanne02}
X.~Chavanne, S.~Balibar, F.~Caupin, C.~Appert, and D.~d'Humi\`eres, J. Low Temp. Phys. \textbf{126}, 643 (2002)

\bibitem{appert01a}
C.Appert, X.~Chavanne, S.~Balibar, D.~d'Humi\`eres, and Ch. Tenaud,
In {\em 4\`emes Rencontres du Non-Lin\'eaire, March 2001},
edited by Uni. Paris~Sud Paris Onze~Editions.

\bibitem{characteristics}
A. Jeffrey and T. Taniuti,
\newblock {\em Non-linear wave propagation},
\newblock (Academic Press, 1964);
\newblock G.B. Whitham,
\newblock {\em Linear and Nonlinear Waves},
\newblock (Wiley-Interscience, 1974).

\bibitem{landau_l6} Landau and Lifshitz,
\newblock in {\em Course of Theoretical Physics, Vol.~6, Fluid Mechanics},
\newblock p.~266 (Pergamon Press, 1959).

\bibitem{caupin98}
F.~Caupin, P.~Roche, S.~Marchand, and S.~Balibar, J. Low Temp. Phys. \textbf{113}, 473 (1998)

\bibitem{jiang_s96}
G.-S. Jiang and C.H. Shu, J. Comput. Physics \textbf{126}, 202 (1996)

\bibitem{tenaud_g_s00}
C.~Tenaud, E.~Garnier, and P.~Sagaut, Int. J. for Numerical Methods in Fluids \textbf{33}, 249 (2000)

\bibitem{shu_o89}
C.W. Shu and S.~Osher, J. Comput. Physics \textbf{83}, 32 (1989)

\bibitem{shu_o88}
C.W. Shu and S.~Osher, J. Comput. Physics \textbf{77}, 439 (1988)

\bibitem{roe81}
P.L. Roe, J. Comput. Physics \textbf{43}, 367 (1981)

\bibitem{chavanne_b_c01c}
X.~Chavanne, S.~Balibar, and F.~Caupin, J. Low Temp. Phys. \textbf{125}, 155 (2001)

\bibitem{chavanne_b_c02}
X.~Chavanne, S.~Balibar, and F.~Caupin, J. Low Temp. Phys. \textbf{126}, 615 (2002)

\bibitem{polar}
R.F.~Harris-Lowe and K.A.~Smee, Phys. Rev. A \textbf{2}, 158 (1970);
\newblock R.J.~Donnelly and C.F.~Barenghi, J. Phys. Chem. Ref. Data \textbf{27}, 1217 (1998);
\newblock C.~Cuthbertson and M.~Cuthbertson, Proc. Roy. Soc. A \textbf{135}, 40 (1932);
\newblock M.H.~Edwards, Can. J. of Phys. \textbf{36}, 884 (1958)

\end{thebibliography}
\end{document}